# Title
Dimensional reduction by geometrical frustration in a cubic antiferromagnet composed of tetrahedral clusters

## Authors
Ryutaro Okuma[1,2,*,@], Maiko Kofu[3], Shinichiro Asai[1], Maxim Avdeev[4,5], Akihiro Koda[6], Hirotaka Okabe[6], Masatoshi Hiraishi[6], Soshi Takeshita[6], Kenji M. Kojima[6,**], Ryosuke Kadono[6], Takatsugu Masuda[1], Kenji Nakajima[3], and Zenji Hiroi[1]

[1]Institute for Solid State Physics, University of Tokyo, Chiba, 277-8581, Japan. [2]Okinawa Institute of Science and Technology Graduate University, Okinawa, 904-0495, Japan. [3]Materials and Life Science Division, J-PARC Center, Japan Atomic Energy Agency, Tokai, Ibaraki, 319-1195, Japan. [4]Australian Nuclear Science and Technology Organization, New Illawarra Road, Lucas Heights, NSW 2234, Australia. [5]School of Chemistry, The University of Sydney, NSW 2006, Australia. [6]Institute of Materials Structure Science, High Energy Accelerator Research Organization (KEK-IMSS), Tsukuba, Ibaraki, 305-0801, Japan.
*Present address: Clarendon Laboratory, University of Oxford, Oxford, OX1 3PU, UK. **Present address: Center for Molecular and Materials Science, TRIUMF, Vancouver, V6T 2A3 Canada.
@e-mail: ryutaro.okuma@gmail.com

## Abstract
Dimensionality is a critical factor in determining the properties of solids and is an apparent built-in character of the crystal structure. However, it can be an emergent and tunable property in geometrically frustrated spin systems. Here, we study the spin dynamics of the tetrahedral cluster antiferromagnet, pharmacosiderite, via muon spin resonance and neutron scattering. We find that the spin correlation exhibits a two-dimensional characteristic despite the isotropic connectivity of tetrahedral clusters made of spin 5/2 $Fe^{3+}$ ions in the three-dimensional cubic crystal, which we ascribe to two-dimensionalisation by geometrical frustration based on spin wave calculations. Moreover, we suggest that even one-dimensionalisation occurs in the decoupled layers, generating low-energy and one-dimensional excitation modes, causing large spin fluctuation in the classical spin system. Pharmacosiderite facilitates studying the emergence of low-dimensionality and manipulating anisotropic responses arising from the dimensionality using an external magnetic field.

## Introduction

All crystalline solids occur in three-dimensional (3D) space but can attain certain quasi-low dimensionalities arising from anisotropic chemical bonds in their crystal structures. For instance, carbon atoms in diamonds possess a 3D network through $sp^3$ bonding, whereas in graphite, they have a two-dimensional (2D) structure through $sp^2$ bonding. Itinerant or localised electrons in a crystal are affected by such anisotropic arrangements of atoms and exhibit a variety of phenomena depending on dimensionality; diamond is hard and insulating, whereas graphite is soft and semimetallic.

Dimensionality critically governs phase transitions and elementary excitations in electronic crystals. In the classical phase transition scenario, a certain symmetry present at high temperatures is spontaneously broken at low temperatures for 3D systems. In contrast, for lower dimensions, according to the Mermin–Wagner theorem, continuous symmetry cannot be spontaneously broken so that the corresponding long-range ordering is prohibited for non-zero temperatures[1]. Hence, low-dimensional systems can realise states of matter that are different from conventional ones and excitations that go beyond the concept of symmetry breaking. Such a system is approximately materialised in an actual 3D crystal via anisotropic chemical bonds: for example, 2D square lattices made of $Cu^{2+}$ ions in cuprate superconductors[2]. Intriguing properties arising from low-dimensional magnetism have been one of the central issues in condensed matter physics.

Low dimensionality in spin systems can be effectively enhanced by geometrical frustration. Let us consider an anisotropic triangular lattice (ATL) consisting of parallel one-dimensional (1D) spin chains with antiferromagnetic (AF) interactions, $J$, which are connected to their neighbours by inter-chain interactions, $J'$, in a zigzag manner (Fig. 1a). In an ATL antiferromagnet, one-dimensionality becomes predominant even with a sizable $J'$ when AF correlations have developed in the chains at low temperatures because the $J'$ couplings are geometrically cancelled out, resulting in a set of decoupled spin chains. This one-dimensionalisation by frustration is theoretically expected even for large $J'$ values up to $J'/J$ = ~0.65[3-5] and has been experimentally observed in $Cs_2CuCl_4$ and $Ca_3ReO_5Cl_2$ comprising ATLs made of spin-1/2 $Cu^{2+}$ and $Re^{6+}$ ions, respectively, with $J'/J$ = 0.3–0.4[6,7]. Moreover, two-dimensionalisation by frustration has been observed for compounds having body-centered tetragonal (BCT) lattices (Fig. 1b) such as $BaCuSi_2O_6$[8,9] and $CePd_2Si_2$[10]. Notably, for these ATL and BCT magnets, the corresponding low dimensionalities are already embedded in the original crystal structures, which create geometrical frustration.

In contrast, magnetic low-dimensionality can emerge purely from a highly frustrated 3D network made of regular tetrahedra (a pyrochlore lattice, Fig. 1c)[11]. In such a system, the global energy minimum is achieved by a null total spin on a tetrahedron, and thus, macroscopically-degenerate spin arrangements. When a small-scale interaction or quantum fluctuation picks up a 3D pattern out of the ensemble, a low-dimensional excitation appears as a result of local spin precession modes. This is observed in the chromium spinel $ZnCr_2O_4$ as a zero-dimensional mode confined in the hexagonal ring of the pyrochlore lattice[12]. 1D and 2D correlations were also observed in spinel $ZnV_2O_4$[13] and double perovskite $Sr_2YRuO_6$[14], respectively. Although certain deformations in the lattice drive the systems to magnetic order and cause low-dimensional excitations in some materials, low dimensionality can, in principle, appear without spatial anisotropy of the underlying lattice.

In this paper, we report a unique 3D tetrahedral-cluster antiferromagnet, pharmacosiderite (($H_3O$)$Fe_4$($AsO_4$)$_3$ ($OH$)$_4$·$5.5H_2O$)[15,16], in which low dimensionality is not present in the crystal structure but emerges purely by geometrical frustration without lattice distortions.

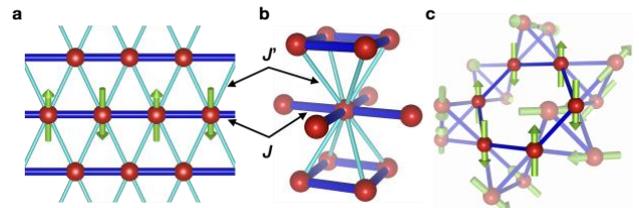

**Fig. 1 | Geometrically frustrated lattices.** The red spheres and green arrows represent magnetic ions and spins, and the thick blue and thin sky-blue sticks represent the strong and

weak magnetic couplings between them, respectively. **a**, Anisotropic triangular lattice composed of 1D chains with spins coupled by AF interaction $J$, which are connected in a staggered manner by $J'$ ($J > J'$). The arrows represent antiferromagnetically aligned spins in the center chain. Two $J'$ interactions in an isosceles triangle cancel each other, effectively eliminating the interchain interactions. **b**, Body-centered tetragonal lattice made of 2D square lattices with $J$ stacked in a staggered manner by $J'$. The four $J'$ couplings are geometrically cancelled when AF correlations develop in the plane. **c**, 3D pyrochlore lattice made of tetrahedra connected by their vertices. The spin arrangement in $ZnCr_2O_4$ contains a cooperative precession mode of six collinear spins on a hexagonal ring, which behaves as a zero-dimensional excitation because interactions with the surrounding antiparallel spins lead to geometrical cancellation[12].

## Results and Discussion

**3D cluster magnet pharmacosiderite.** Pharmacosiderite with the general formula $AFe_4(AsO_4)_3(OH)_4 \cdot nH_2O$ (A is a monovalent cation) crystallises into a cubic structure comprising clusters made of four $FeO_6$ octahedra (Fig. 2a, Supplementary Note 1). The magnetic sublattice consists of regular tetrahedral clusters made of spin-5/2 $Fe^{3+}$ ions arranged in a primitive cubic cell (Fig. 2b); a related lattice with tetrahedra arranged in a face-centred cubic lattice is the breathing pyrochlore lattice[17]. The intra- and intercluster magnetic interactions, $J$ and $J'$, are estimated to be AF with magnitudes of 10.6 and 2.9 K, respectively, from the fitting of the magnetic susceptibility (Fig. 3a)[16]. Notably, there is a strong frustration in both the tetrahedral cluster and the intercluster couplings because the four $J'$ paths form an elongated tetrahedron along the cube edges.

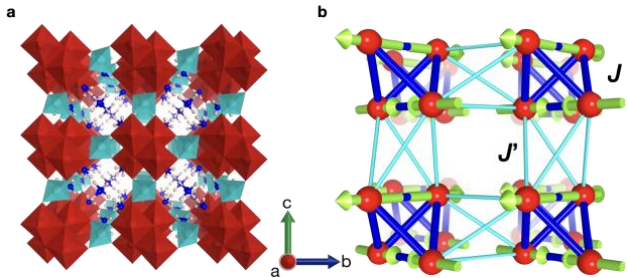

**Fig. 2** | Crystal and magnetic structures of pharmacosiderite, $(D_3O)Fe_4(AsO_4)_3(OD)_4 \cdot 5.5D_2O$. **a**, Cubic crystal structure of the space group $P\bar{4}3m$. The red octahedra, sky-blue tetrahedra, blue spheres, and pink spheres represent $FeO_6$ octahedral units, $AsO_4$ tetrahedral units, oxygens, and hydrogens of water molecules, respectively. The red, blue, and green arrows indicate the crystallographic $a$, $b$, and $c$ axes, respectively. Four $FeO_6$ octahedra with small trigonal distortions are connected by their edges to form a cluster of $T_d$ symmetry, forming a regular tetrahedron made of $Fe^{3+}$ ions. These clusters are located at the vertices of the cubic unit cell with ~8 Å on each edge and connected to neighbours via a $[AsO_4]^{3-}$ tetrahedron along the edge of the cube. The cube possesses a large open space at the centre, which accommodates various large monovalent ions, a hydronium ion in this case, and water molecules. **b**, Fe sublattice and the $\mathbf{q} = 0$, $\Gamma_5$ magnetic structure below $T_N = 6$ K. Red spheres represent spin-5/2 $Fe^{3+}$ ions, four of which form a regular tetrahedral cluster at each vertex of the cubic unit cell. The yellow green arrows represent the directions of the magnetic moments of Fe spins, all of which lie approximately in the planes perpendicular to the $c$ axis. The thick blue and thin sky-blue sticks represent intracluster Heisenberg interactions $J$ and intercluster Heisenberg interactions $J'$, respectively.

A magnetic long-range order (LRO) sets in below $T_N = 6$ K with a weak ferromagnetic moment of $7 \times 10^{-3}$ $\mu_B$ predominantly along the [100] axis (Fig. 3a). Interestingly, however, previous Mössbauer spectroscopy measurements depicted an broad spectrum even in the LRO state (Fig. 3b)[16]. Such broadening could arise from structural inhomogeneity (Supplementary Note 1). However, it is more likely that the broad spectrum is predominantly caused by the phenomenon in which the ordered spin is constantly flipped at a frequency of ~$10^2$ MHz (Supplementary Note 2). Indeed, our muon spin relaxation (µSR) experiments confirmed such a dynamical fluctuation with a similar time scale in the ordered state, as revealed by Mössbauer spectroscopy (Fig. 3c and Supplementary Note 3). The implanted muon experiences a fluctuating local field overlaid on a static internal field, which is evidenced by the observation that the muon spin exhibits an exponential relaxation of spin polarisation by the fluctuating internal field after an initial recovery due to the static field even under longitudinal fields up to 0.4 T. These results strongly suggest the coexistence of two types of magnetism with different dynamics at the microscopic level. Hence, the LRO may be unconventional, and an intriguing phenomenon is awaiting discovery.

To obtain a deeper insight into the magnetism of pharmacosiderite, we carried out neutron scattering experiments on a polycrystalline sample of $(D_3O)Fe_4(AsO_4)_3(OD)_4 \cdot 5.5D_2O$. The magnetic properties of pharmacosiderite are insensitive to the choice of the A-site cation (Supplementary Note 3). Figure 3d shows the magnetic contribution at 1.6 K; a nuclear contribution has been subtracted using the 10 K data as a reference. No structural transition was observed down to 1.6 K (Supplementary Note 1); hence, the cubic symmetry was preserved, which is in agreement with previous X-ray diffraction experiments[16]. The magnetic peaks have nearly the same widths as the nuclear peaks (Supplementary Note 4), indicative of a long-range magnetic order; the magnetic correlation length estimated is 360 Å, which is 45 times larger than the $a$-axis length.

All the magnetic Bragg peaks appeared in the same positions as the nuclear peaks, indicating a $\mathbf{q} = 0$ magnetic order. Thus, spin arrangements in a single tetrahedron are sufficient to be considered. Because the minimum requirement is to make the total spin zero in each tetrahedron, there are five possible spin arrangements (Supplementary Note 4). Our Rietveld fitting to the data pinned down a $\mathbf{q} = 0$, $\Gamma_5$ magnetic structure, as depicted in Fig. 2b.

The spin arrangement of the $\mathbf{q} = 0$, $\Gamma_5$ magnetic order is coplanar in the (001) plane with two pairs of antiparallel spins in one tetrahedron aligned along the two edges perpendicular to the $c$ axis; it is a type of two-in, two-out structure with local [110] anisotropy and not [111] anisotropy as in the spin-ice compounds[18]. A small canting of each spin is possible along the $c$ axis, resulting in a minuscule net moment when a weak single-ion anisotropy exists along the local [111] direction. Notably, the magnetic structure is essentially tetragonal, but we observed no tetragonal lattice distortion, probably because of negligible spin–lattice couplings owing to the indirect structural connection between clusters via the $AsO_4$ units, which is different from the case of spinel compounds[11].

A notable finding of the magnetic structure analysis is that the magnitude of the refined magnetic moment at 1.6 K is 2.08(4) $\mu_B$, which is much smaller than the expected value of 5 $\mu_B$ for a

spin-5/2 system; the ordered moment is almost saturated at this temperature, judging from the temperature evolution of the (100) peak intensity (Supplementary Figure 10c). This significant reduction in the ordered magnetic moment is consistent with the persistent fluctuations observed in the μSR and Mössbauer spectroscopy. However, we should note that the large nuclear $R_{wp}$ factor of ~27 %, which mostly comes from partial deuteration, suggests that the standard deviation of the ordered moment value derived from the Rietveld analysis is somewhat underestimated (Supplementary Note 1).

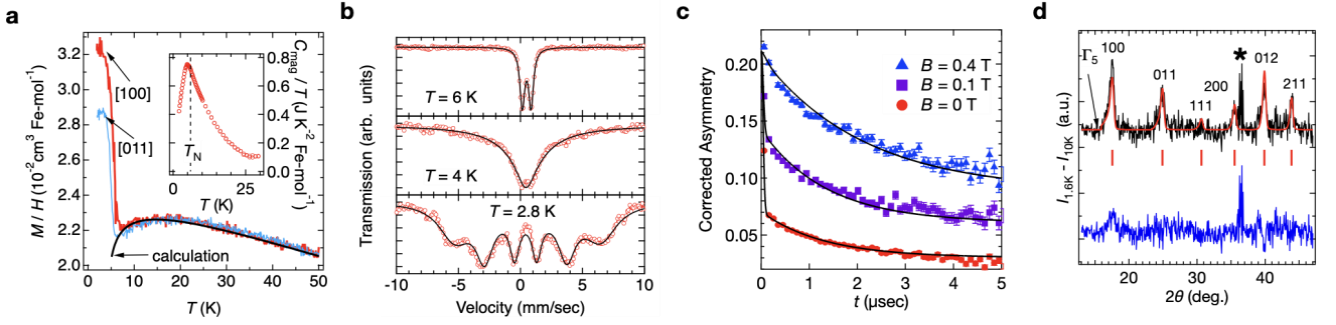

**Fig. 3 | Experimental data demonstrating an LRO in pharmacosiderite. a**, Temperature dependencies of the magnetization $M$ divided by magnetic field $H$ measured under magnetic fields of $\mu_0 H$ = 0.1 T applied along the [100] and [011] directions of a single crystal of pharmacosiderite (main panel) and magnetic heat capacity $C_{mag}/T$ at zero field (Inset). The curve on the $M/H$ data represents a fit to a coupled-cluster model, which yields $J$ = 10.6 K and $J'$ = 2.9 K[16]. **b**, Mössbauer spectra of the powder sample of pharmacosiderite at $T$ = 2.8, 4, and 6 K[16]. The black line is a fit to a Blume–Tjon model[19] taking into account fluctuating local magnetic fields (Supplementary Note 2). Even below $T_N$, the spectra are broad, and the intensity ratio deviates from the conventional powder average, suggesting magnetic fluctuations remaining in the LRO. **c**, Time-dependent $\mu^+$–$e^+$ decay asymmetry, which is proportional to the muon spin polarization, measured at 2 K in a zero external field and longitudinal fields of 0.1 and 0.4 T, shown by red circle, purple square, and blue triangle, respectively. The error bars indicate the standard deviations of the muon decay asymmetry at each point. The curve on each dataset represents a fit to a model in which one muon site near a hydrogen feels a static local field and the other muon site feels both static and fluctuating local fields (Supplementary Note 3). **d**, Powder neutron diffraction pattern and a Rietveld fitting to the **q** = 0, $\Gamma_5$ spin structure depicted in Fig. 2b. The black line represents data taken at 1.6 K after the subtraction of the 10 K data as a reference of nuclear contributions. The red line, blue line, and green bars represent a Rietveld fit, a residual of fitting, and the positions of magnetic Bragg peaks, respectively. The asterisk * indicates the residual from the subtraction of the 10 K data from the 1.6 K.

**Inelastic neutron scattering experiments.** To obtain information on the magnetic excitations and the origin of the spin fluctuations, we performed inelastic neutron scattering experiments. A powder-averaged dynamical structure factor $S(Q, E)$ obtained at an incident neutron energy of $E_i$ = 7.7 meV at $T$ = 1.6 K is shown in Fig. 4a. Large intensities are observed at approximately $Q/\text{Å}^{-1}$ = 0.8 and 1.6, which correspond to the (100) and (200) reflections in the elastic channel, respectively. Dispersive modes develop from these and form domes with maxima at ~4 meV, which are apparently spin wave (SW) excitations from the **q** = 0 magnetic order. In contrast, an $S(Q, E)$ obtained at an $E_i$ of 3.1 meV and $T$ = 0.6 K, which is more sensitive to low-energy excitations, evidences a small gap with a magnitude of 0.5 meV (Fig. 4b and Supplementary Note 5).

The observed $S(Q, E)$s were compared with simulations based on the linear SW theory for the **q** = 0 , $\Gamma_5$ structure after convolution with the instrumental resolution (Fig. 4c). The overall features, the bandwidth and intensity distribution, are well reproduced by the calculation with no adjustable parameters; $J$ = 0.9 meV and $J'$ = 0.27 meV from the fitting of magnetic susceptibility data[16]. Small intracluster interactions were added to account for the gap in the spectrum; a Dzyaloshinskii–Moriya (DM) interaction of 0.01$J$, which is likely to exist and contributes to selecting the **q** = 0 and $\Gamma_5$ order among others, as well as an easy-axis single ion anisotropy of 0.001$J$, which causes a small spin canting, are necessary to explain the weak ferromagnetism observed by magnetometry. However, the observed spectra are much broader than the calculated ones, especially along the energy axis (Supplementary Note 6), which is consistent with the reduced ordered moment in the ground state.

To investigate the spin fluctuations in pharmacosiderite, we examined the $S(Q, E)$ data in terms of the equal-time structure factor of spin correlation perpendicular to a scattering vector $\langle S_i^\perp(0) \cdot S_j^\perp(0)\rangle$, which corresponds to a Fourier transform of a spin configurations snapshot. Because most of the intensity is concentrated between 0 and 2 meV, as shown in Supplementary Figs. 11 and 12, the total scattering is approximated by integrating $S(Q, E)$ within this energy range as follows:

$$I(Q) \propto \int_0^{2\ \text{meV}} S(Q,E) dE \sim \int_{-\infty}^{\infty} S(Q,E) dE$$
$$= \sum_{ij} e^{iQ(r_i - r_j)} \langle S_i^\perp \cdot S_j^\perp \rangle \quad (1).$$

The temperature evolution of $I(Q)$ is plotted in Fig. 4d. Broad peaks indicative of a short-range order (SRO) are observed above $T_N$, and additional sharp Bragg peaks from the LRO appear below $T_N$, with the broad peaks almost unchanged. The large intensity of the broad peaks relative to those of the Bragg peaks indicates the presence of strong spin fluctuations in pharmacosiderite even in the LRO state.

A striking feature of the broad peaks is their asymmetric shape. Such a peak shape with a long tail toward high $Q$ is generally known to be a characteristic of the powder diffraction profile from low-dimensional spin systems. In this study, we analysed the paramagnetic scattering of pharmacosiderite at 7.2 K in terms of the low-dimensional scattering model[20,21] (Supplementary Note 7). As shown in Fig. 4e, the 2D model best reproduces the experimental spectrum with a coherence length within the plane $2\pi D^{-1}$ estimated to be 9 times the lattice constant, ~70 Å. Thus, a 2D SRO exists in pharmacosiderite despite the cubic symmetry of the underlying lattice. Because similar broad peaks remain below $T_N$, this 2D SRO must survive as a large 2D spin fluctuation in the LRO.

By employing the linear SW theory for the Heisenberg spin model, we have examined whether such a 2D spin fluctuation is, in fact, possible in pharmacosiderite. Figure 4f shows the calculated SW dispersions for the **q** = 0, $\Gamma_5$ magnetic structure. Four SW modes appear as there are four sites in the unit cell. Notably, all of them are completely flat at zero energy along $\Gamma$–Z without DM interactions (solid lines in Fig. 4f) or only weakly dispersive at finite energies with DM interactions (dash-dotted lines), while being highly dispersive along most other directions. This strongly suggests that the motion of magnons tends to be confined within the $c$ layer in the low-energy region, leading to two-dimensionality in the spin excitation despite the isotropic magnetic couplings.

**Two-dimensionalisation by frustration.** Here, we discuss the mechanism leading to the observed two-dimensionality in pharmacosiderite within the classical Heisenberg spin model. Figure 5a illustrates the intercluster couplings in the **q** = 0 and $\Gamma_5$ structures with the $c$ plane as a coplanar plane. A spin in a tetrahedron interacts with three pairs of antiferromagnetically aligned spins in the nearby tetrahedra via the identical Heisenberg interaction $J'$. Notably, towards the pair along the $c$ axis, the two $J'$ paths (sky-blue bonds in Fig. 5a) are geometrically cancelled, whereas the $J'$ paths (orange bonds) toward either pair along the $a$ or $b$ axis do not cause such a cancellation. As a result, the 3D network is transformed into an array of decoupled layers with the remaining in-plane interactions of the order of $J'$. The 2D feature in the $S(Q, E)$s observed in our experiments, and the SW calculations may be attributed to this dimensional reduction or two-dimensionalisation by geometrical frustration. We emphasise that, compared to the ATL or BCT antiferromagnets (Fig. 1), the resulting 2D anisotropy is not a feature of the original crystal lattice but is induced by the evolution of AF correlations. Because the unique axis perpendicular to the layers is not fixed to a certain crystallographic direction, it should be equally chosen among all three identical [100] directions. Further, it can be controlled by external magnetic fields, as will be mentioned later.

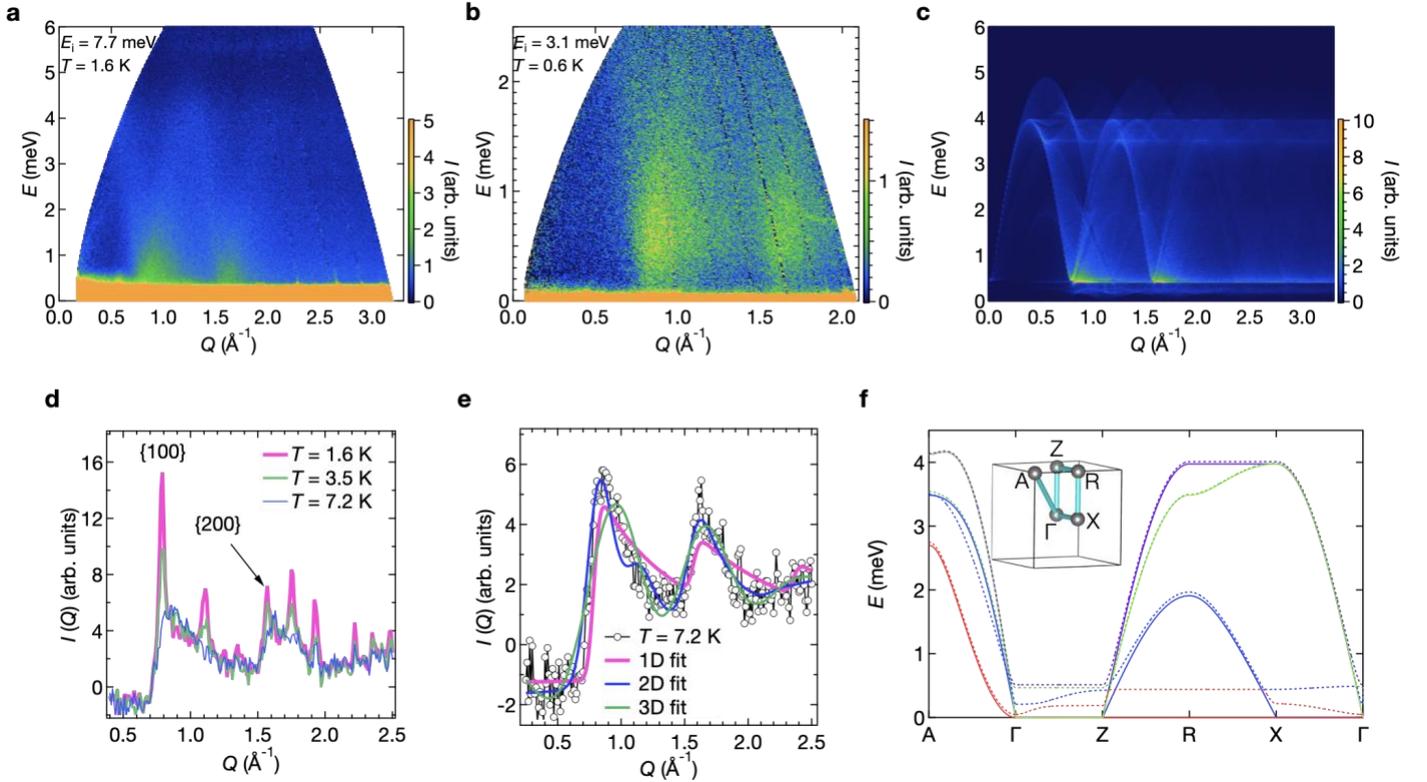

**Fig. 4 | Inelastic neutron scattering experiments from a powder sample of pharmacosiderite and simulations based on the linear SW theory. a**, **b**, Powder-averaged equal-time structure factors $S(Q, E)$ measured at $T$ = 1.6 K and an incident neutron energy of $E_i$ = 7.7 meV (**a**) and at $T$ = 0.6 K and $E_i$ = 3.1 meV (**b**). The colour scales of the neutron intensity are shown on the right side of **c**. **c**, Simulated powder-averaged $S(Q, E)$ calculated by the linear SW theory for the **q** = 0, $\Gamma_5$ magnetic structure. The isotropic interactions $J$ = 0.9 meV and $J'$ = 0.27 meV, which were determined from magnetic susceptibility measurements[16], are assumed. In addition, an intracluster DM interaction of $0.01J$ and an easy-axis single ion anisotropy of $0.001J$ along [111] are included in the calculation. **d**, Temperature evolution of the integrated structure factor $I(q)$ deduced from the $S(Q, E)$ after the subtraction of the data collected at 50 K as a reference for the lattice contribution (see text). The magenta, green, and blue lines represent the data obtained at 1.6, 3.5, and 7.2 K, respectively. **e**, Fitting of the $I(q)$ at 7.2 K to the low-dimensional scattering model. The magenta, green, and blue lines represent fits for 1D, 2D, and 3D scattering, respectively. The agreement factors $R_{wp}$ are 6.0, 3.3, and 6.6 %, respectively, indicating that the 2D model best describes the data. See Supplementary Note 7 for details of the fits. **f**, Dispersion relation of magnetic excitations calculated for a single crystal. SWs between the selected high symmetry $k$ points in the Brillouin zone shown in the inset are calculated by the linear SW theory for the **q** = 0, $\Gamma_5$ magnetic structure with the same parameters used for **c**. The solid and dash-dotted lines represent the dispersions without and with anisotropic interactions, respectively. As there are four sites in the magnetic unit cell, four modes distinguished by different colours appear.

In an actual crystal of pharmacosiderite, we must consider a modification arising from small anisotropic interactions in addition to the major Heisenberg interactions. The geometrical cancellation in the $J'$ couplings along the $c$ axis should become imperfect owing to the DM interaction and the spin canting by the single ion anisotropy, as reflected by the appearance of the weak dispersions along Γ–Z in the SW calculations (Fig. 4f). Nevertheless, as these anisotropy terms are quite small, to preserve a certain flatness in the Γ–Z modes, the influence may be negligible except at sufficiently low temperatures; judging from the dispersions in 0–0.2 meV, two-dimensionalisation likely occurs above ~2 K.

We have demonstrated that 2D spin fluctuations exist in the **q** = 0, Γ$_5$ order of pharmacosiderite. In general, however, such a 2D fluctuation can reduce the ordered magnetic moment slightly, but not that much. In fact, our SW calculations with the anisotropies found a reduction to 4.0 $\mu_B$ at 0 K, which is still much larger than the experimental value of 2.1 $\mu_B$. However, this calculated value is close to the maximum static magnetic moment estimated by Mössbauer spectroscopy. This observation means that the local magnetic moment itself is not reduced by quantum fluctuations, whereas the average one over the LRO is likely reduced by certain spatial and temporal fluctuations.

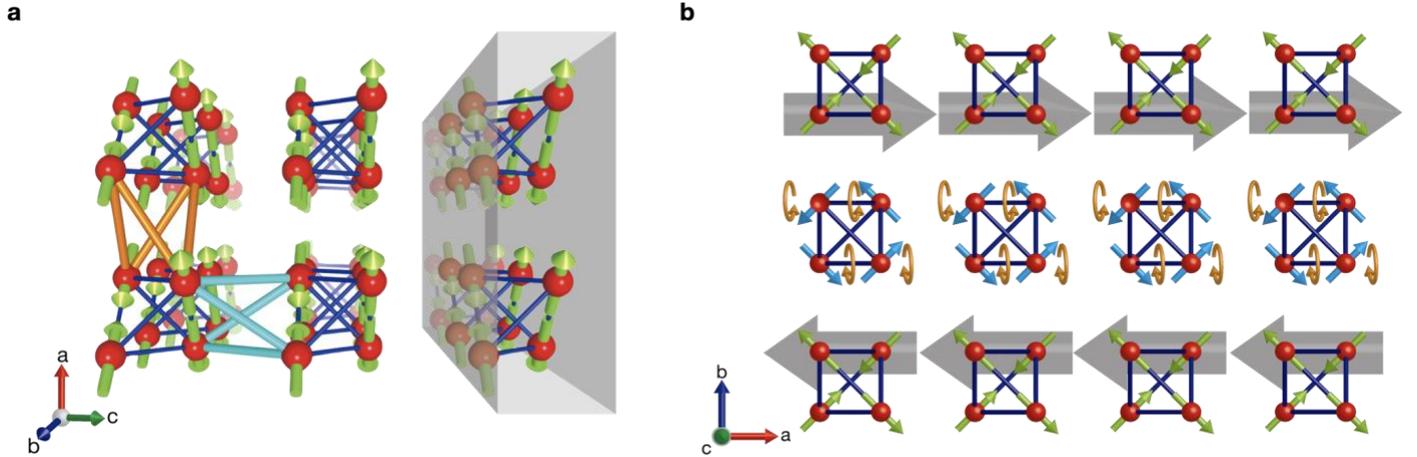

**Fig. 5 | Two- and one-dimensionalisation by geometrical frustration in pharmacosiderite. a,** Schematic representation of the **q** = 0, Γ$_5$ magnetic structure with all the spins perpendicular to the horizontal $c$ axis. The red, blue, and green arrows indicate the crystallographic $a$, $b$, and $c$ axes, respectively. Among the six intercluster $J'$ paths from a focused spin in a tetrahedron, the two paths to the pair of antiparallel spins along the $c$ axis (sky-blue bonds) geometrically cancel each other because the total $S$ equals 0 for the pair. In contrast, the paths to either pair of spins along the $a$ or $b$ axis (orange bonds) do not; spins in either pair are not antiparallel to each other. This results in a slab perpendicular to the $c$ axis, in which SWs are effectively confined. **b,** 1D defect along the $a$ axis, which is made of Γ$_4$ tetrahedra with blue spins and is generated without energy loss in the slab shown in **a** of the **q** = 0 magnetic structure made of Γ$_5$ tetrahedra with green spins. The red, blue, and green arrows indicate the crystallographic $a$, $b$, and $c$ axes, respectively. The large grey arrow represents the sum of a pair of spins at an edge of the tetrahedron next to the 1D defect, which is parallel to the $a$ axis and exerts an effective axial magnetic field to the spins of the 1D defect. Under the axial effective fields, all the blue Γ$_4$ spins in the 1D defect can rotate collectively along the $a$ axis while keeping the total spin zero and are continuously transformed into the Γ$_5$ arrangement without losing energy in the intercluster couplings to the surrounding. This means that such a 1D defect can be generated and annihilated without energy loss in the 2D slab of the **q** = 0, Γ$_5$ magnetic structure in the classical spin model.

**One-dimensionalisation generating 1D defects**. The origin of the large reduction in the ordered moment may be ascribed to the 1D spin fluctuations generated in the 2D layer of the **q** = 0, Γ$_5$ order. Notably, in the SW calculations of Fig. 4f without anisotropies, there is a zero-energy, flat excitation mode along Γ–Z–R–X–Γ (red curve), in addition to those along Γ–Z mentioned above. This mode is not a local zero-dimensional mode because it exhibits a large dispersion along Γ–Z. Instead, it may be a 1D mode occurring along the $a$ axis because its dispersion is flat along the $b^*c^*$ plane. We plausibly assume that the 1D mode originates from a line defect in the Γ$_5$ layer as a result of geometrical cancellation of intercluster interactions in the isotropic spin model. As depicted in Fig. 5b and described in detail in Supplementary Note 8, all the spins in a string of clusters along the $a$ axis can rotate collectively so as to keep all the intercluster couplings to the surroundings always antiferromagnetic. As a result, there is no energy cost to generate such a 1D defect made of Γ$_4$ clusters in the **q** = 0, Γ$_5$ order as a zero-energy mode. Therefore, one-dimensionalisation by frustration should take place together with two-dimensionalisation in this unique spin system. More solid experimental evidence may require further inelastic neutron scattering studies using a single crystal of pharmacosiderite.

True zero-energy excitation modes should destroy the LRO completely. However, the order-by-disorder effect may lift the degeneracy in the presence of thermal or quantum fluctuation[22,23]. In the present compound, the small anisotropies also lift it to have a finite energy gap; this is not exactly a true energy gap but a pseudo gap with a few dispersive modes existing below the flat modes in the SW calculation in Fig. 4f. Thus, they are scarcely populated at $T = 0$. Nevertheless, at high temperatures, many of them can be thermally activated and partly destroy the LRO, which may produce a large 1D spin fluctuation that causes a reduction in the average ordered moment in the LRO. This may be the reason for the observed large reduction in the ordered moment and the coexistence of the static and dynamic magnetisms probed by Mössbauer and μSR experiments (Supplementary Note 2 and 3).

**Possible thermal conductivity switching**. Finally, we propose a concept based on the present findings in pharmacosiderite. It is well established that an array of spins can carry a heat flow in addition to the lattice[24]. Specifically, in 1D spin systems, the thermal conductivity is notably enhanced along the chains[25,26]. Spin contributions to thermal conductivity can be larger than lattice contributions in some frustrated magnets[27-29]. For pharmacosiderite, one expects a larger thermal conductivity along the $c$ layer as magnon excitations are confined in the layer. Interestingly, the direction of the $c$-axis can be controlled by an external magnetic field because it is not fixed to the crystallographic lattice. Rather, it is selected by the direction of the weak ferromagnetic moment of the magnetic order; even a small magnetic field of 0.5 T can control it[16]. Thus, the thermal conductivity along any direction is enhanced and reduced when the field is perpendicular and parallel to the direction, respectively, as schematically illustrated in Fig. 6. Notably, a difference in thermal conductivity along the two directions only measures the spin contributions. Therefore, magnetic-field switching of thermal conductivity might be possible in pharmacosiderite. Future thermal conductivity measurements using a single crystal or even a polycrystalline sample would confirm this intriguing idea.

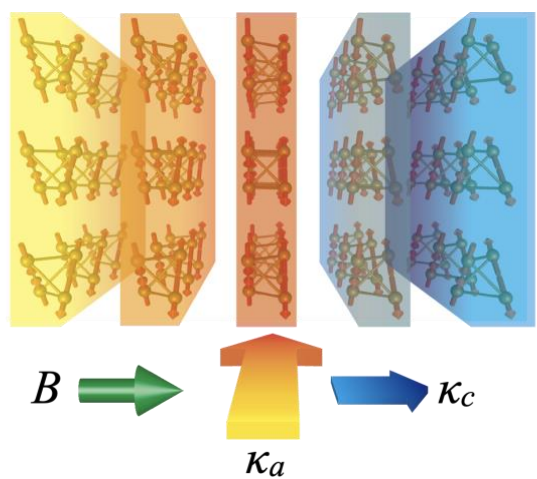

**Fig. 6 | Concept of magnetic-field control of thermal conductivity.** When magnetic field $B$ is applied horizontally along one of the cubic axes of pharmacosiderite in its $q = 0$, $\Gamma_5$ magnetic order, magnon excitations are confined in each vertically aligned layer; this selection of the magnetic $c$ axis occurs because the parasitic ferromagnetic moment along the $c$ axis prefers the field direction even if $B$ is as small as 0.5 T. Because magnons can carry a heat flow, the thermal conductivities $\kappa_a$ and $\kappa_c$ become large and small along directions perpendicular and parallel to $B$, respectively, which results in a large horizontal temperature gradient, as depicted by the variation in colour. When $B$ is rotated to the vertical direction, the anisotropic thermal conduction is reversed. Thus, a switching of thermal conductance becomes possible via the external magnetic field.

## Methods

**Sample preparation.** A powder sample of pharmacosiderite was synthesised by a conventional hydrothermal method[16]. First, a beige colloidal precursor was obtained by vigorous stirring of a ferric solution (30 g of $NH_4Fe(SO_4)_2 \cdot 12H_2O$ fully dissolved in 7.5 g of water) and an arsenate solution (8 g of $KH_2AsO_4$ and 7.5 g of $K_2CO_3$ fully dissolved in 19 g of water). After the pH of the precursor was adjusted to approximately 1.5 by the addition of 0.1 mL of 10M HClaq, the precursor was heated in a Teflon-lined autoclave for three h at 220 °C. After the reaction, a pale-yellow precipitate of potassium pharmacosiderite $KFe_4(AsO_4)_3(OH)_4 \cdot nH_2O$ (n = 8–9) was filtered and thoroughly washed with water. Subsequently, the obtained powder was annealed in 500 mL of 0.1M HClaq for a week at 100 °C. Finally, hydronium pharmacosiderite $(H_3O)Fe_4(AsO_4)_3(OH)_4 \cdot 5.5H_2O$ was obtained as a pale-green powder after filtration and drying. Deuterated ingredients were used for the neutron scattering experiments. Cation and deuterium substitution have negligible effects on the magnetic properties. Single crystals of pharmacosiderite were synthesised under hydrothermal conditions under a temperature gradient[30]. A quartz ampoule with a length of 150 mm was filled with 0.1 g of polycrystalline pharmacosiderite and 5 mL of solution remained after powder synthesis. A thick quartz tube with outer and inner diameters of 12 and 8 mm, respectively, was used to avoid explosion. The ampoule was placed vertically in a two-zone furnace and heated at 200 °C at the hot bottom end and 120 °C at the cold top end for a month. Seething of the solution caused polycrystalline pharmacosiderite to decompose and recrystallise near the liquid level. This resulted in the growth of cube-shaped crystals as large as 0.3×0.3×0.3 mm.

**Magnetization measurements.**
Magnetisation measurements were conducted using a single crystal of pharmacosiderite in a magnetic property measurement system 3 (MPMS-3, Quantum Design).

**Heat capacity measurements.**
Heat capacity measurements were conducted using powder pharmacosiderite in a physical property measurement system (Quantum Design).

**Muon spin rotation/relaxation measurements** Conventional μSR experiments were conducted on two types of pharmacosiderite: $AFe_4(AsO_4)_3(OH)_4 \cdot nH_2O$ with A = K and $H_3O$. The K sample was examined in an ARTEMIS spectrometer in the S1 area at the J-PARC MUSE. A 100% spin-polarized pulsed beam of positive muons with a momentum of 27 MeV/c and a full-width-at-half-maximum of $t_{PW} \simeq 100$ ns was delivered to a powder sample loaded on a He-flow cryostat. Time-dependent μ–e decay asymmetry $A(t)$ was measured under a zero field (ZF), a longitudinal field (LF), or a weak transverse field (TF) over a temperature range from 4 to 300 K. Additional measurements for the $H_3O$ sample were conducted in a Lampf spectrometer on the M20 beamline at TRIUMF, Canada, to observe a fast depolarisation emerging at 2 K in the early time range of $A(t)$ with a higher time resolution of ∼1 ns.

**Neutron diffraction experiments** Powder neutron diffraction experiments were performed using constant-wavelength (λ = 2.4395 Å) diffractometer ECHIDNA in ANSTO at 1.6 and 10 K, which are above and below $T_N = 6$ K. Rietveld refinements were performed to determine the magnetic structure using the program Fullprof Suite[31]. Crystal and magnetic structures were drawn using a VESTA[32].

**Inelastic neutron scattering experiments** Inelastic neutron scattering experiments were performed in a time-of-flight spectrometer AMATERAS[33] at J-PARC at 0.6, 1.6, 2.5, 3.5, 5, 7.2, 8, 20, and 50 K at incident energies of 3.1 meV and 7.7 meV, where the energy resolutions at the elastic position were 0.058 and 0.43 meV (full width at half maximum), respectively. The raw data were reduced by the UTSUSEMI software suite[34].

**Spin-wave calculations.**
Linear SW analysis was conducted using the SpinW software[35].

## Data Availability
The publication data used in this study is available at https://doi.org/10.6084/m9.figshare.13996436.v1. The data files are also available from the corresponding author upon request.

# Acknowledgments

We are grateful to Gøran. J. Nilsen, H. Yan, H. Tsunetsugu, M. Ogata, N. Shannon, S. Hayashida, Y. Kato and Y. Motome for helpful discussions. We thank Kazuaki Iwasa for providing us with a cryostat for inelastic neutron scattering experiments. R.O. is supported by the Materials Education Programme for the Future Leaders in Research, Industry, and Technology (MERIT) given by the Ministry of Education, Culture, Sports, Science and Technology of Japan (MEXT). This work was partially supported by KAKENHI (Grant No. 15K17701) and the Core-to-Core Programme for Advanced Research Networks given by the Japan Society for the Promotion of Science (JSPS). The neutron diffraction experiment at ECHIDNA was supported by a General User Program for Neutron Scattering Experiments, Institute for Solid State Physics, The University of Tokyo (Proposal No. 17401 and No. 17403). The neutron and muon experiment at the Materials and Life Science Experimental Facility of the J-PARC was performed under approved proposals No. 2016MI21, 2017I0014, and 2018I0014.

## Author contributions

R.O., M.K., and Z.H. conceived and designed the study. R.O., M.K., and K.N. performed inelastic neutron scattering experiments. S.A., M.A., and T.M. performed the diffraction experiments. S.A. analysed the diffraction data. A.K., H. O., M. H., S. T., K. M. K., and R.K. performed the μSR experiments. A.K. analysed the μSR data. R.O., M.K., and Z.H. interpreted all the experimental data. R.O. performed linear SW calculations. The manuscript was written based on the discussion by all authors.

## Competing interests

The authors declare no competing interests.

# Supplementary Information for "Dimensional reduction by geometrical frustration in a cubic antiferromagnet composed of tetrahedral clusters"

Ryutaro Okuma[1,2*@], Maiko Kofu[3], Shinichiro Asai[1], Maxim Avdeev[4,5], Akihiro Koda[6], Hirotaka Okabe[6], Masatoshi Hiraishi[6], Soshi Takeshita[6], Kenji M. Kojima[6**], Ryosuke Kadono[6], Takatsugu Masuda[1], Kenji Nakajima[3], and Zenji Hiroi[1]

[1]Institute for Solid State Physics, University of Tokyo, Chiba, 277-8581, Japan. [2]Okinawa Institute of Science and Technology Graduate University, Okinawa, 904-0495, Japan. [3]Materials and Life Science Division, J-PARC Center, Japan Atomic Energy Agency, Tokai, Ibaraki, 319-1195, Japan. [4]Australian Nuclear Science and Technology Organization, New Illawarra Road, Lucas Heights, NSW 2234, Australia. [5]School of Chemistry, The University of Sydney, NSW 2006, Australia. [6]Institute of Materials Structure Science, High Energy Accelerator Research Organization (KEK-IMSS), Tsukuba, Ibaraki, 305-0801, Japan.
*Present address: Clarendon Laboratory, University of Oxford, Oxford, OX1 3PU, UK. **Present address: Center for Molecular and Materials Science, TRIUMF, Vancouver, V6T 2A3 Canada.
@e-mail: ryutaro.okuma@gmail.com

**Supplementary Note 1:   CRYSTAL STRUCTURE REFINEMENTS OF PHARMACOSIDERITE**

A nuclear structural refinement of a deuterated pharmacosiderite, $(D_3O)Fe_4(OD)_4(AsO_4)_3 \cdot 5.5D_2O$, was performed for powder neutron diffraction data using the Fullprof Suite[1]. Because the sample contained hexagonal $D_2O$ ice as a minor phase, a two-phase fit was performed, as shown in Supplementary Fig. 1a, which yielded the structural parameters listed in Supplementary Table 1. The initial structure models employed were based on the parameters deduced from the previous powder X-ray diffraction (XRD) analysis[2] for pharmacosiderite and the hexagonal structure of water ice $I_h$ for $D_2O$ ice[3]. According to ref. 16, Fe1, As1, O1, and O2 are fully occupied, forming a framework with cavities where O3 and O4 statistically reside forming water molecules or hydronium ions. Thanks to the large neutron cross section of deuterium, deuterium positions in these units were also determined, which was difficult to achieve in the previous XRD study. D2 of the hydroxy group (Supplementary Fig. 1b) is fully ordered and located at a 4$e$ site, similar to O2. Concerning the deuterium atoms related to water, namely, D1 and D3, they inevitably possess orientational disorders because of the low point group symmetry of a water molecule. D1 and D3 are placed at 12$i$ positions with fractional occupancies around the O3 and O4 atoms, respectively. The occupancy of D1 and O3 is fixed at half such that the water molecule made from both sites must take one of two orientations around the face-centre position of the unit cell. The trigonal pyramidal unit made of O4 and D3 is compatible with a hydronium ion ($D_3O^+$) or a disordered water molecule. The fractional occupancies of O4 and D3 were determined to fulfil the chemical formula at 87.5 %. A soft constraint over the bond length between the D and O atoms was applied to all deuterium atoms. The fitting appears reasonably good in Supplementary Fig. 1a, but the agreement factor $R_{wp}$ in the final refinement remains as high as 26.5 %. This is probably due to incomplete deuteration and large occupational disorder at the O3, O4, D1, and D3 sites.

Two kinds of crystal water or hydronium molecules in the framework are highlighted in Supplementary Figs. 1c and 1d, and their influence on magnetic interactions was considered. Water molecules made of D1 and O3 must have little effect on magnetic interactions because they are distant from Fe and not involved in the magnetic paths. In contrast, water molecules or hydronium ions made of D3 and O4 must affect the magnetic path more seriously because they form a hydrogen bond with D2, which is a part of the hydroxide ligand of Fe. Thus, the orientational and occupational disorder of the D3–O4 unit may moderate the intracluster interaction to some extent but with little influence on the intercluster interaction.

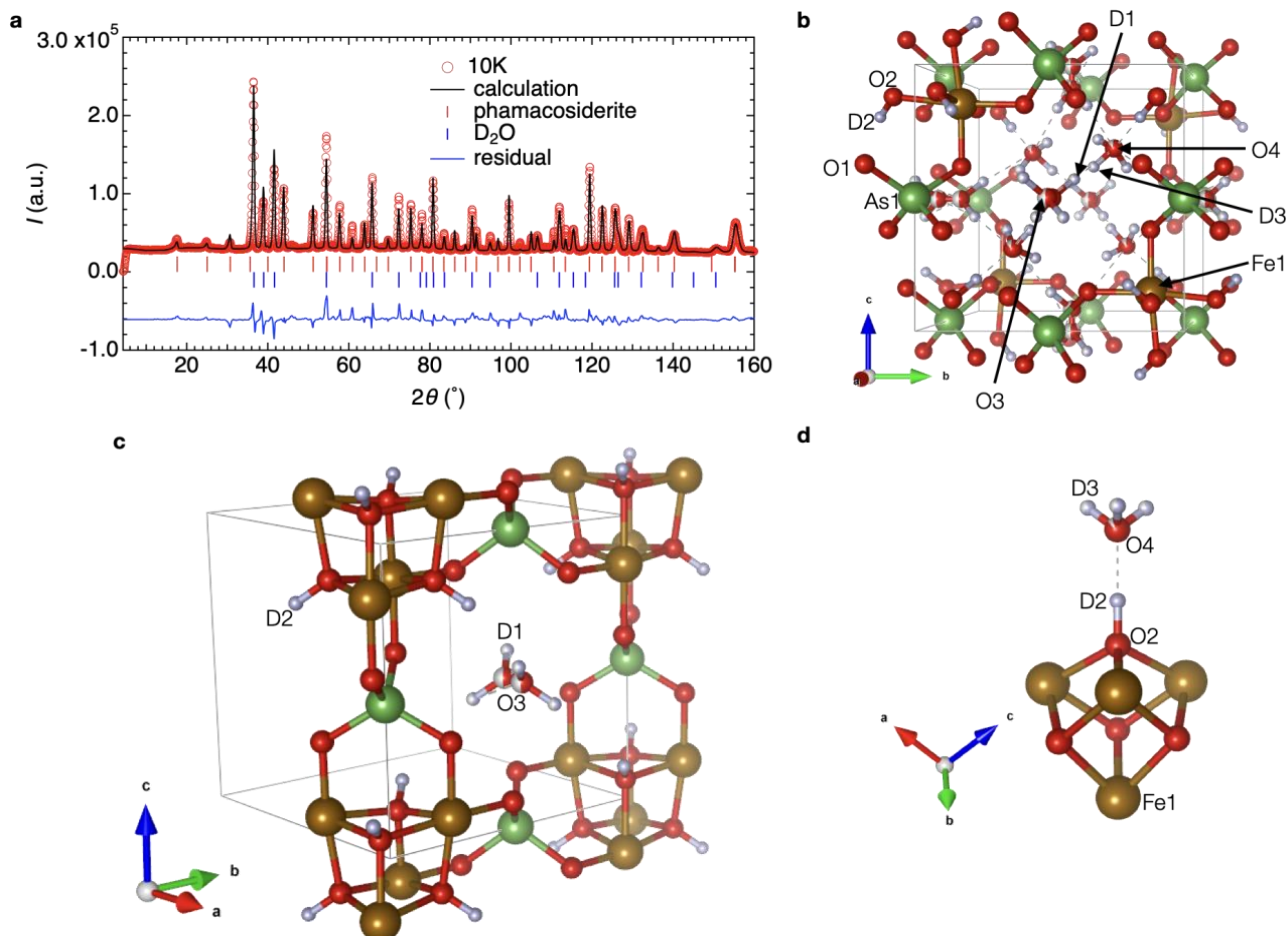

**Supplementary Fig. 1 | Nuclear structural refinement on the powder neutron diffraction data. a.** Red marks, the black line, and blue line represent raw data at 10K, a fit, and a residual, respectively. The red and blue bars indicate the positions of the Bragg peaks of pharmacosiderite and ice, respectively. **b**. Refined crystal structure of pharmacosiderite. **c**. Disorder of a water molecule near the face centred position in pharmacosiderite. **c**. Disorder of a water molecule or hydronium ion near a hydroxy group in pharmacosiderite.

| | Formula | | | $(D_3O)Fe_4(OD)_4(AsO_4)_3 \cdot 5.5D_2O$ | | |
|---|---|---|---|---|---|---|
| | Space group | | | $P\text{-}43m$ | | |
| | $a$ / Å | | | 7.99822(8) | | |
| | $V$ / Å$^3$ | | | 511.659(9) | | |
| | $Z$ | | | 1 | | |
| Atom | Wyckoff position | $x$ | $y$ | $z$ | $B_{iso}$ | Occupancy |
| Fe | 4$e$ | 0.13972(63) | 0.13972(63) | 0.13972(63) | 1 | 1 |
| As | 3$d$ | 0.5 | 0 | 0 | 1 | 1 |
| O1 | 12$i$ | 0.14129(68) | 0.14129(68) | 0.37936(85) | 1 | 1 |
| O2 | 4$e$ | 0.88733(53) | 0.88733(53) | 0.88733(53) | 1 | 1 |
| O3 | 6$g$ | 0.5 | 0.04543(156) | 0.5 | 1 | 0.5 |
| O4 | 4$e$ | 0.70465(73) | 0.70465(73) | 0.70465(73) | 1 | 0.875 |
| D1 | 12$i$ | 0.57478(102) | 0.11877(183) | 0.42527(102) | 3 | 0.5 |
| D2 | 4$e$ | 0.81455(51) | 0.81455(51) | 0.81455(51) | 3 | 1 |
| D3 | 12$i$ | 0.62774(69) | 0.62774(69) | 0.73458(109) | 3 | 0.667 |

| Fomula | $D_2O$ |
|---|---|

| Space group | | | $P6_3/mmc$ | | |
|---|---|---|---|---|---|
| $a$ / Å | | | 4.50072(9) | | |
| $c$ / Å | | | 7.33031(24) | | |
| $V$ / Å³ | | | 128.593(6) | | |
| $Z$ | | | 1 | | |
| Atom | $x$ | $y$ | $z$ | $B_{iso}$ | Occ. |
| O | 1/3 | 2/3 | 0.05986(81) | 1 | 1 |
| D1 | 0.45797(68) | 0.91595(136) | 0.02377(69) | 3 | 1/2 |
| D2 | 1/3 | 2/3 | 0.18954(114) | 3 | 1/2 |

**Supplementary Table 1 | Structural parameters of pharmacosiderite and D₂O $I_h$ ice obtained via the Rietveld refinements.**

**Supplementary Note 2: ANALYSIS OF MÖSSBAUER SPECTRA**

The low-temperature Mössbauer spectra of pharmacosiderite were analysed based on the Blume–Tjon model[4]. It is assumed in the model that a local magnetic field from spins takes either $+h$ or $-h$ with probabilities per unit time $W_1$ and $W_2$, respectively, and flips between the two states randomly. In a magnetically ordered state, $W = (W_1 + W_2)/2$ and $\alpha = (W_1 - W_2)/(W_1 + W_2)$ correspond to the average frequency of magnetic fluctuation and the degree of magnetic ordering, respectively[5]. The angle made by the principal axis of the electric field gradient (three-fold axis in pharmacosiderite) and the local field is designated as $\theta$. The results of fitting are good, as shown in Fig. 2b. The refined parameters are presented in Supplementary Table 2. The values of the isomer shift (IS) and quadrupole splitting (QS) determined by fitting the paramagnetic data at 6 K were used for the analyses of the 4 and 2.8 K data. The peak width of the underlying Lorentzian function $\Gamma$ was refined for each dataset considering variable inhomogeneity. The data at 4 K were analysed as paramagnetic because the refined $\alpha$ value was negligible.

| $T$ (K) | $\Gamma$ (mm/sec) | IS (mm/sec) | QS (mm/sec) | $h$ (T) | $W$ (MHz) | $\alpha$ (%) | $\theta$ (°) |
|---|---|---|---|---|---|---|---|
| 6 | 0.441(5) | 0.494(2) | 0.1176(6) | 0 | – | – | – |
| 4 | 1.42(9) | 0.494 | 0.1176 | 61 | 254(23) | 0 | 49(2) |
| 2.8 | 0.86(3) | 0.494 | 0.1176 | 61(5) | 313(72) | 57(5) | 45(1) |

**Supplementary Table 2 | Mössbauer parameters of pharmacosiderite. Fixed parameters in the fitting are given without uncertainty.**

The obtained average local frequency of magnetic fluctuation is approximately 300 MHz, which is the same order as that detected by μSR in Section 3. The direction of the local field is 45(1)° at 2.8 K, which is roughly consistent with the magnetic structure determined by neutron diffraction in Section 4: $\theta$ takes $\cos^{-1}(2/\sqrt{6}) \sim 35°$ in the $\Gamma_5$ structure (Supplementary Fig. 2).

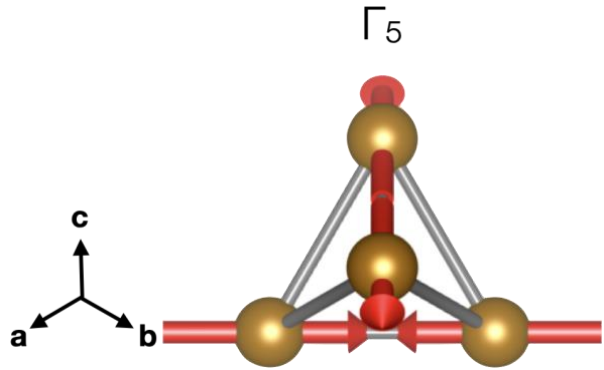

**Supplementary Fig. 2 | Local spin arrangement viewed along the [111] direction, which is the principal axis of the electric field gradient at an Fe site.** In the $\Gamma_5$ magnetic structure two of the four spins are perpendicular to the viewing direction, and the other 2 spins make an angle of ~35º to the viewing direction.

**Supplementary Note 3: LOCAL SPIN DYNAMICS PROBED BY μSR**

Supplementary Fig. 3 shows the time evolution of the positron decay asymmetry [$A(t)$, which we call the time spectrum] in the paramagnetic phase of K-pharmacosiderite [KFe₄(AsO₄)₃(OH)₄·nH₂O]. Here, we used a pharmacosiderite with a K ion inside the cage instead of a hydronium ion; the magnetic properties were found to be insensitive to the type of cage ion. In general, $A(t)$ is proportional to the instantaneous muon polarisation projected along the $z$ axis parallel to the initial muon beam direction; $A(t) = A_0 P_z(t)$, where $A_0$ is the instrumental asymmetry and is equal to ~0.23. The time spectrum at 59 K well above $T_N$ under a zero external field (ZF) is characterised by a slow exponential damping that overlaps with a sinusoidal oscillation. This oscillatory signal are attributed to neither magnetic order nor muonium formation, the former of which is excluded by the fact that $A(0)$ exhibits full polarisation. The origin of the oscillation may be attributed to the formation of a local atomic cluster consisting of a muon and a small number (≤ 2) of nearby atoms having nuclear magnetic moments. One such classical example is the F$^-$–$\mu^+$–F$^-$ (F$\mu$F$^-$) complex observed in alkali metal fluorides, where two farad nuclei (spin $I = 1/2$) and one muon form a local three-spin system that gives rise to coherent spin precession[6]; the F$\mu$F$^-$ state is a muonic analogue of hydrogen difluoride (FHF$^-$), which is known as a prototype of strong hydrogen bonding[7]. Another example is found in sodium alanate, NaAlH$_4$, where the alanate ion [AlH$_4$]$^-$ substitutes the role of fluorine in the interstitial muon to form an H$\mu$H-bonding state[8]. These examples indicate a common tendency for a muon to behave as a pseudo-hydrogen and form a hydrogen bond with negatively charged light ions. Thus, we presume that a part of the implanted muons in pharmacosiderite also form a similar local spin complex with the H of the hydroxyl bases, giving the observed sinusoidal oscillation.

Assuming a static collinear geometry with $\mu^+$ at the centre of the line joining the two other nuclear spins ($I = 1/2$), the time evolution of muon polarisation as a cubic average is calculated by solving a simple three-spin model to yield

$$G_{3S}(t) = \frac{1}{6}[3 + \cos(\sqrt{3}\omega_d t) + \alpha_+\cos(\beta_+\omega_d t) + \alpha_-\cos(\beta_-\omega_d t)] \quad (1),$$

where $\alpha_\pm = 1 \pm 1/\sqrt{3}$, $\beta_\pm = (3 \pm \sqrt{3})/2$, and $\omega_d$ is the dipolar interaction frequency.

$$\omega_d = 2\gamma_\mu \gamma_I / r^3 \quad (2),$$

where $\gamma_I$ and $r$ are the gyromagnetic ratios of nuclear spin ($\gamma_I/2\pi = 42.58$ MHz/T for $^1$H) and the distance between $\mu^+$ and the nucleus. On the other hand, it is also known that a muon may occupy a site bonding with one of hydroxyl bases[9, 10], leading to a two-spin model given by the following function:

$$G_{2S}(t) = \frac{1}{6}\left[1 + \cos(\omega_d t) + 2\cos\left(\frac{1}{2}\omega_d t\right) + 2\cos\left(\frac{3}{2}\omega_d t\right)\right] \quad (3).$$

The time spectrum under a longitudinal field of 0.05 T exhibits exponential damping, whereas the spin depolarisation due to the nuclear dipolar field is quenched in the high field limit ($B_0 \gg \omega_d/\gamma_\mu \sim 10^{-3}$ T). This indicates that a considerable fraction of implanted muons are subjected to exponential depolarisation due to fluctuation of local fields exerted by paramagnetic moments.

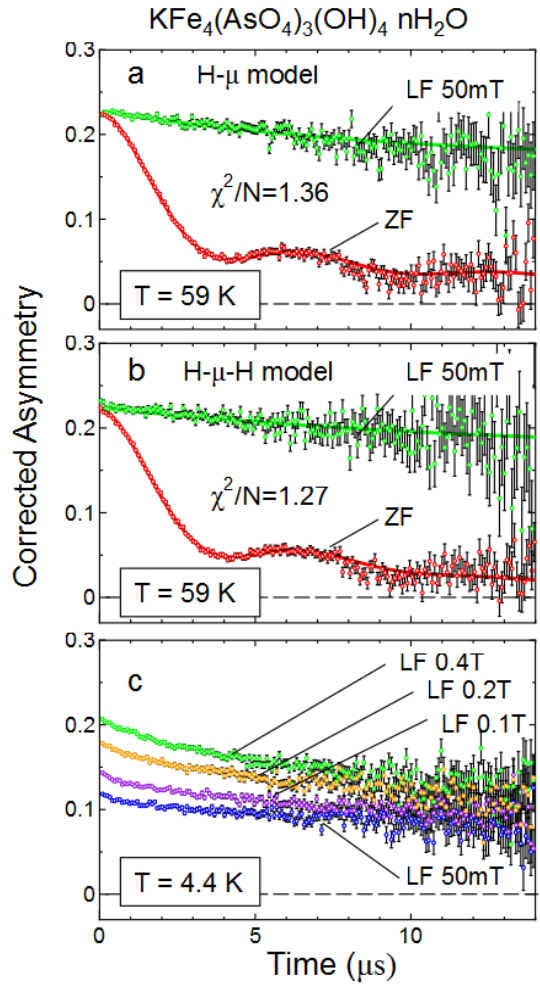

**Supplementary Fig. 3 | a, b. Typical μSR time spectra observed in K-pharmacosiderite at 59 K under ZF and longitudinal fields (LF) of 0.05 T.** A clear sinusoidal oscillation is seen under ZF, which is readily quenched upon applying a small LF of 0.05 T. The relaxing signal is analysed by assuming an H–μ bonding (**a**) and an H–μ–H bonding (**b**) (see text). Besides, the remaining

exponential damping under LF is attributed to the fluctuating Fe moments. **c**, Spectra observed at 4.4 K below $T_N$ showing a recovery of $A(0)$ with increasing longitudinal fields, which indicates that the fast depolarization exceeding the time resolution of muon pulses of ~0.1 μs is induced by a quasi-static internal field.

The time spectra below $T_N$, shown in Supplementary Fig. 3c, exhibits a recovery of $A(0)$ with increasing longitudinal fields, indicating that a part of the signal undergoes a fast depolarisation induced by a quasi-static local field; a similar feature is observed for $H_3O$-pharmacosiderite, as shown in Fig. 3c in the main text. The time evolution for this component under LF = $B_0$ ($x = B_0/B_{loc}$) is approximately given by

$$G_z^{stat}(t,x) \simeq G_\infty(x) + [1 - G_\infty(x)]\exp(-\sigma_s^2 t^2) \quad (4),$$

$$G_\infty(x) = \frac{3x^2 - 1}{4x^2} + \frac{(x^2-1)^2}{16x^3}\log\left[\frac{(x+1)^2}{(x-1)^2}\right] \quad (5),$$

where $G_\infty(x)$ corresponds to the powder average of the probability that the net local field $\mathbf{B}_{loc} + \mathbf{B}_0$ is parallel to the initial muon spin polarisation, and $\sigma_s$ (> 1 μs$^{-1}$) denotes the relaxation rate. In addition, a remaining component exhibits an exponential depolarisation under LF, which may be described by

$$G_z^{dyn}(t,\omega_\mu) \simeq \exp[-\Lambda_d(\omega_\mu)t] \quad (6),$$

$$\Lambda_d(\omega_\mu) \simeq \frac{2\delta_\mu^2 \nu}{\omega_\mu^2 + \nu^2} \quad (7),$$

where $\omega_\mu = \gamma_\mu B_0$, $\delta_\mu$ is the muon–Fe hyperfine field, and $\nu$ is the fluctuation rate of $\delta_\mu$. Notably, Supplementary Eq. (7) becomes least dependent on $B_0$ for $\nu \gg \omega_\mu$, which mimics the behaviour of the exponentially depolarizing component for $A(t)$.

Considering these three components, the time spectra in Supplementary Fig. 3 are analysed using a variety of models and are best reproduced by the following model relaxation functions:

$$A(t) = A_0\left[(1-f)G_{3S}(t) + fG_z^{dyn}(t,\omega_\mu)\right] (T > T_N) \quad (8),$$

$$A(t) = A_0\left[(1-f)G_{3S}(t) + fG_z^{dyn}(t,\omega_\mu)\right]G_z^{stat}(t,x) (T < T_N) \quad (9),$$

where $f$ is the fractional yield of the component exhibiting persistent dynamical modulation. The parameters deduced from the fitting analysis are summarised in Supplementary Table 3. The fitting analysis using $G_{2S}(t)$ instead of $G_{3S}(t)$ gives a slightly larger value of $\chi^2/N$, as shown in Supplementary Figs. 3a and 3b. In addition, the use of $G_{3S}(t)G_z^{dyn}(t,\omega_\mu)$ instead of $G_{3S}(t)$ does not improve the fitting. Thus, we use the model given by Supplementary Eqs. (8) and (9) in the following discussion. It is plausible that the observed two magnetic sectors, (1 - $f$) and $f$, come from different muon sites, which are referred to μ(1) and μ(2), respectively. We note that the parameter $f$ is common in both analyses below and above $T_N$.

|  | H-μ model ($G_{2S}(t)$) | H-μ-H model ($G_{3S}(t)$) |
|---|---|---|
| $f$ | 0.30 | 0.21 |
| $\omega_d / 2\pi$ [MHz] | 0.10189(8) | 0.06645(6) |
| $r$ [nm] | 0.1554(4) | 0.1792(5) |
| $\chi^2/N$ | 1.36 | 1.27 |

**Supplementary Table 3 | Summary of the fitting results assuming both cases of an H-μ bonding and an H-μ-H bonding.**

According to the above model, a muon at the μ(1) site forms an H–μ–H bonding state with nearby H atoms, and probes a quasi-static internal field $\mathbf{B}_{loc}$ from the ordered Fe moments below $T_N$. The H–μ bond length is estimated to be 0.1792(5) nm from the magnitude of $\omega_d$ (= $2\pi \times 0.06645(6)$ MHz) deduced by the curve fit. On the other hand, a muon at the μ(2) site is exposed to an additional local field that possesses a persistent fluctuation, as indicated by the fact that a better fit is obtained by assuming a product of $G_z^{dyn}(t,\omega_\mu)$ and $G_z^{stat}(t,x)$ below $T_N$. Thus, the μ(2) probes two kinds of Fe moments with different dynamics: one is quasi-static and the other is fluctuating. This leads to the speculation that the magnetic order is somewhat inhomogeneous. The persistent fluctuating field at the μ(2) site for $H_3O$-pharmacosiderite estimated by the analysis of the $B_0$ dependence of $\Lambda_d$ at 2 K using Supplementary Eq. (7) is $\nu \simeq 4.27(10) \times 10^8$ s$^{-1}$ with $\delta_\mu \simeq 12.8(2)$ MHz, indicating a very small energy scale of 2 μeV, which is orders of magnitude smaller than that observed by the inelastic neutron scattering experiments. Thus, the μSR experiments detect much slower spin fluctuations in the LRO of the Fe moments, which must be different from the 2D spin fluctuations found by inelastic neutron scattering experiments. It is plausible to assume that the slow spin fluctuations come from the 1D defect generated in the decoupled 2D layers, as inferred by the SW calculations. The μ(2) site is located near the 1D defect and is subject to both quasi-static and fluctuating local fields. The latter is induced by a collective mode in the 1D defect, as shown in Fig. 5b in the main text, whereas the μ(1) site is located far from the 1D defect.

**Supplementary Note 4: MAGNETIC STRUCTURE ANALYSIS**

The spin configuration of the **q** = 0 magnetic structure in pharmacosiderite was analysed based on the representation analysis using SARAh suite[11]. The $Fe^{3+}$ atom resides at a 4$e$ site with $x \sim 0.14$ in the space group of $P\bar{4}3m$; 4 inequivalent sites in the unit cell are labelled as 1: ($x, x, x$), 2: ($x, -x, -x$), 3: ($-x, x, -x$), 4: ($-x, -x, x$). The magnetic representation for a $q = 0$ structure can be decomposed into irreducible representations (IR) as follows:

$$\Gamma_q = \Gamma_2 + 2\Gamma_2 + 3\Gamma_4 + 3\Gamma_5^2$$

Supplementary Table 4 lists twelve possible basis vectors (BV), $\psi_1, \cdots, \psi_{12}$, with the components of ($m_a$, $m_b$, $m_c$) for each of the four sites. Supplementary Fig. 4 illustrates all five types of spin configurations with zero net moment for the tetrahedron. Powder neutron diffraction intensities are calculated for these magnetic structures, as shown in Supplementary Fig. 5. Notably, there are intense (100) and (200) peaks in the experimental pattern of Fig. 3d, which are reproduced only for $\Gamma_4$, $\psi_4$ and $\Gamma_5$, $\psi_7 - \psi_8$. Thus, we have carried out magnetic Rietveld refinements for the two magnetic structures, as shown in Supplementary Fig. 6. The agreement factor is significantly smaller for the $\Gamma_5$ structure than for the $\Gamma_4$ structure: $R_{wp} = 29.8$ % and 44.7 %, respectively. This fact indicates that the $\Gamma_5$ structure is realized in pharmacosiderite. In addition, a small net moment along [001] is consistent with $\Gamma_5$ structure, which is not allowed for $\Gamma_4$[16].

As shown in Supplementary Fig. 7, the magnetic order is long-range in nature, which is evident from the similar peak widths of the nuclear and magnetic (001) reflections. Estimation of the magnetic correlation length by microstructural analysis yielded a magnetic correlation length of 360.7(3) Å.

| IR | BV | Site | $m_a$ | $m_b$ | $m_c$ |
|---|---|---|---|---|---|
| $\Gamma_2$ | $\psi_1$ | 1 | 0.5774 | 0.5774 | 0.5774 |
|  |  | 2 | 0.5774 | -0.5774 | -0.5774 |
|  |  | 3 | -0.5774 | 0.5774 | -0.5774 |
|  |  | 4 | -0.5774 | -0.5774 | 0.5774 |
| $\Gamma_3$ | $\psi_2$ | 1 | 0.7071 | -0.7071 | 0 |
|  |  | 2 | 0.7071 | 0.7071 | 0 |
|  |  | 3 | -0.7071 | -0.7071 | 0 |
|  |  | 4 | -0.7071 | 0.7071 | 0 |
|  | $\psi_3$ | 1 | 0.4082 | 0.4082 | -0.8165 |
|  |  | 2 | 0.4082 | -0.4082 | 0.8165 |
|  |  | 3 | -0.4082 | 0.4082 | 0.8165 |
|  |  | 4 | -0.4082 | -0.4082 | -0.8165 |
| $\Gamma_4$ | $\psi_4$ | 1 | 0 | -0.7071 | 0.7071 |
|  |  | 2 | 0 | 0.7071 | -0.7071 |
|  |  | 3 | 0 | 0.7071 | 0.7071 |
|  |  | 4 | 0 | -0.7071 | -0.7071 |
|  | $\psi_5$ | 1 | 0.7071 | 0 | -0.7071 |
|  |  | 2 | -0.7071 | 0 | -0.7071 |
|  |  | 3 | -0.7071 | 0 | 0.7071 |
|  |  | 4 | 0.7071 | 0 | 0.7071 |
|  | $\psi_6$ | 1 | -0.7071 | 0.7071 | 0 |
|  |  | 2 | 0.7071 | 0.7071 | 0 |
|  |  | 3 | -0.7071 | -0.7071 | 0 |
|  |  | 4 | 0.7071 | -0.7071 | 0 |
| $\Gamma_5$ | $\psi_7 - \psi_8$ | 1 | 0 | 0.7071 | 0.7071 |
|  |  | 2 | 0 | -0.7071 | -0.7071 |
|  |  | 3 | 0 | -0.7071 | 0.7071 |
|  |  | 4 | 0 | 0.7071 | -0.7071 |
|  | $\psi_7 + 2\psi_8$ | 1 | 1 | 0 | 0 |
|  |  | 2 | 1 | 0 | 0 |
|  |  | 3 | 1 | 0 | 0 |
|  |  | 4 | 1 | 0 | 0 |
|  | $\psi_9 - \psi_{10}$ | 1 | 0.7071 | 0 | 0.7071 |
| | | 2 | -0.7071 | 0 | 0.7071 |
| | | 3 | -0.7071 | 0 | -0.7071 |
| | | 4 | 0.7071 | 0 | -0.7071 |
| | $\psi_9 + 2\psi_{10}$ | 1 | 0 | 1 | 0 |
| | | 2 | 0 | 1 | 0 |
| | | 3 | 0 | 1 | 0 |
| | | 4 | 0 | 1 | 0 |
| | $\psi_{11} - \psi_{12}$ | 1 | 0.7071 | 0.7071 | 0 |
| | | 2 | -0.7071 | 0.7071 | 0 |
| | | 3 | 0.7071 | -0.7071 | 0 |
| | | 4 | -0.7071 | -0.7071 | 0 |
| | $\psi_{11} + 2\psi_{12}$ | 1 | 0 | 0 | 1 |
| | | 2 | 0 | 0 | 1 |
| | | 3 | 0 | 0 | 1 |
| | | 4 | 0 | 0 | 1 |

**Supplementary Table 4 | Irreducible representation (IR) and the components ($m_a$, $m_b$, $m_c$) of basis vectors (BV) for the magnetic representation in the space group $P\bar{4}3m$ and the magnetic propagation vector q = 0.**

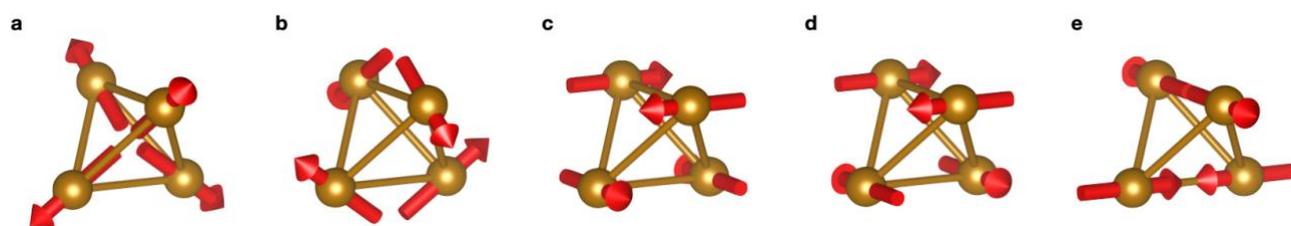

**Supplementary Fig. 4 | Representative basis vectors of irreducible representations for the q = 0 magnetic structure in pharmacosiderite. a.** $\Gamma_2$, $\psi_1$. **b.** $\Gamma_3$, $\psi_2$. **c.** $\Gamma_3$, $\psi_3$. **d.** $\Gamma_4$, $\psi_4$. **e.** $\Gamma_5$, $\psi_7 - \psi_8$.

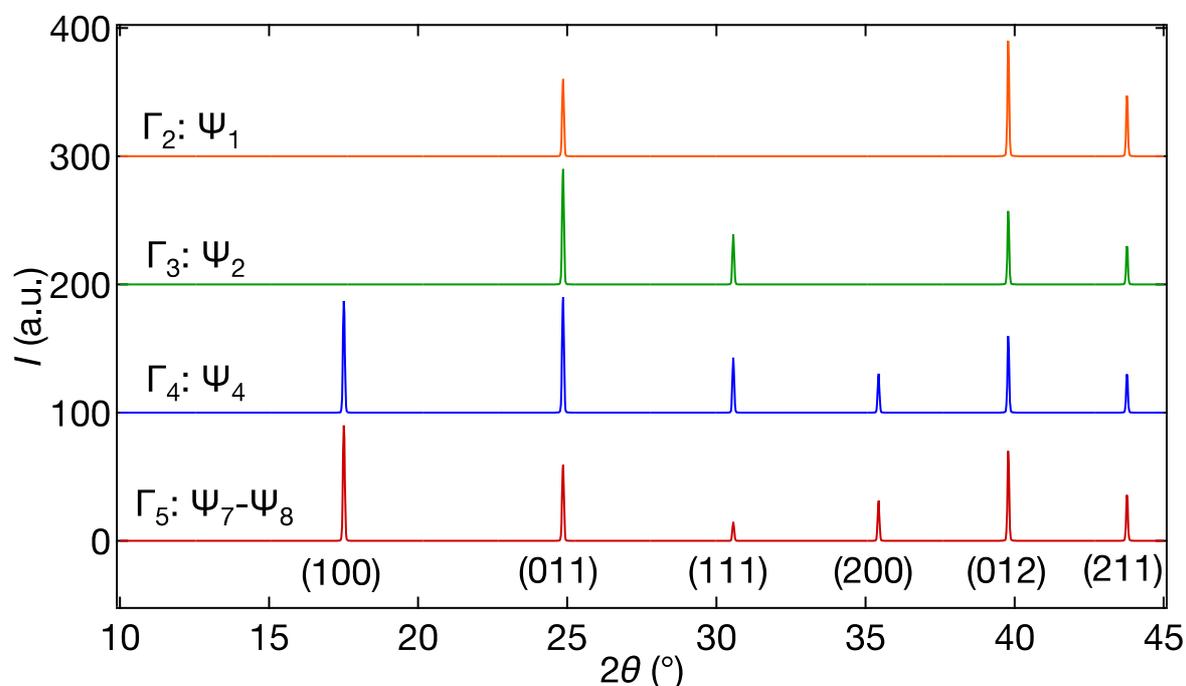

**Supplementary Fig. 5 | Simulated powder neutron diffraction intensities for the q = 0 magnetic structures shown in Supplementary Fig. 4.** The simulation for $\psi_3$ is omitted since it is same as that of $\psi_2$.

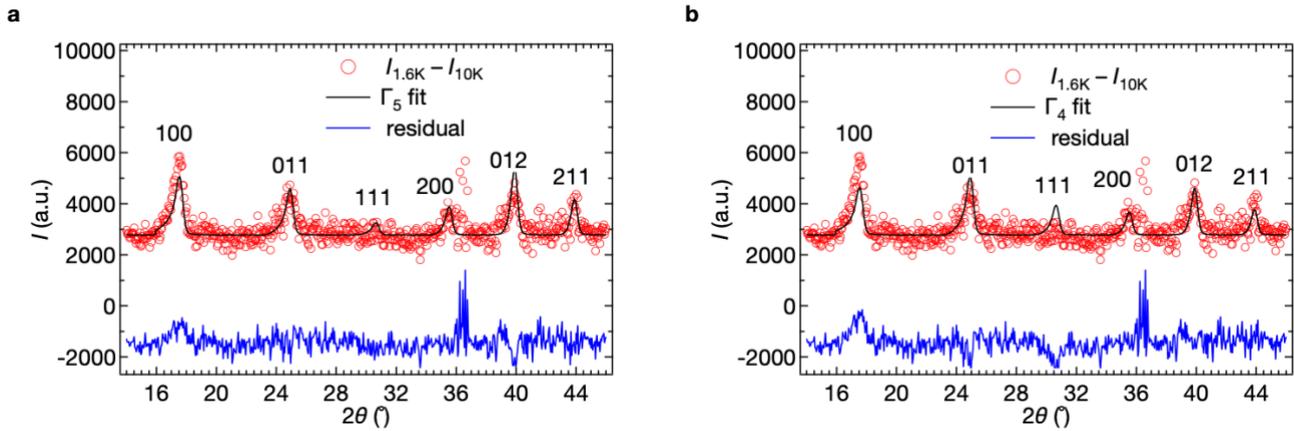

**Supplementary Fig. 6 | Powder neutron diffraction pattern of pharmacosiderite and its Rietveld fittings to a. $\Gamma_4$ and $\psi_4$ and b. $\Gamma_5$ and $\psi_7 - \psi_8$ spin structures.** The red circles represent data taken at 1.6 K after the subtraction of the 10 K data as a reference of nuclear contributions. The black and blue lines represent a fit and a residual in each figure.

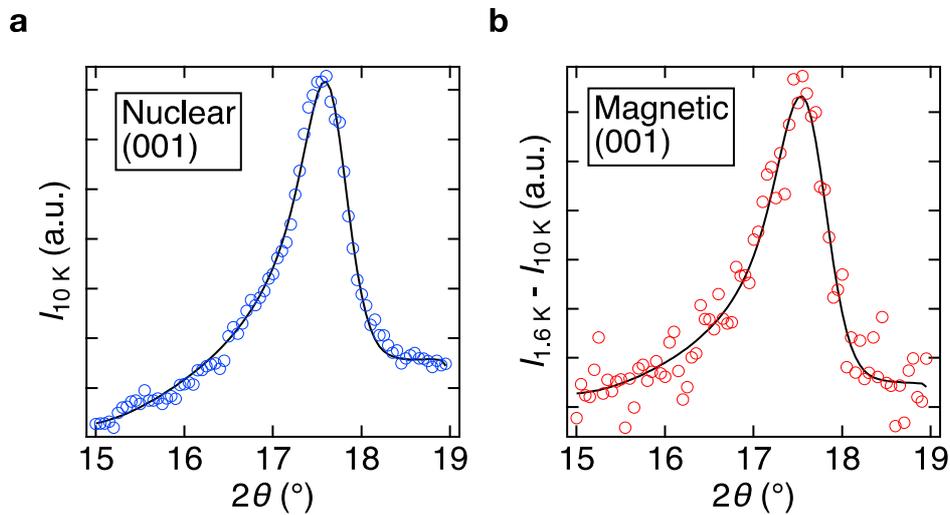

**Supplementary Fig. 7 | Comparison of (001) peak in the a. nuclear and b. magnetic diffraction.** The blue and red circles represent data taken at 10 K and data taken at 1.6 K after the subtraction of the 10 K data, respectively. The black lines represent fits to the Thompson–Cox–Hastings pseudo-Voigt function convoluted with the axial divergence asymmetry function installed in Fullprof[31]. FWHMs of the magnetic and nuclear peaks are 0.51 and 0.57 degrees, respectively.

**Supplementary Note 5:    TEMPERATURE DEPENDENCE OF THE DYNAMICAL STRUCTURE FACTOR**

The temperature evolution of the powder-averaged dynamical structure factor $S(Q, E)$ at $E_i = 3.1$ meV is shown in Supplementary Fig. 8. To estimate the energy dependence of the intensity, $S(Q, E)$ is integrated along $Q$ around the (001) reflection (Supplementary Fig. 9). The intense elastic component is located within ±0.1 meV (Supplementary Fig. 9a). An opening of a gap of ~1 meV is clearly observed upon cooling at the tail of the peak in Supplementary Fig. 9b. To estimate the magnetic diffraction pattern, $S(Q, E)$ is integrated at an energy window between –0.1 and 0.1 meV (Supplementary Fig. 10a). The magnetic profile shown in the inset of Supplementary Fig. 10a resembles that of Fig. 3d in the elastic channel and is consistent with the $q = 0$, $\Gamma_5$ structure. The temperature dependence of the (001) intensity is shown in Supplementary Fig. 10c, which is obtained by a Gaussian fit to each (001) peak shown in Supplementary Fig. 10b. The intensity increases below the Neél temperature $T_N = 6$ K and almost saturates below 2.5 K. Thus, the magnetic long-range order has already developed at 0.6 K with the ordered magnetic moment almost saturated at this temperature.

The temperature evolution of $S(Q, \omega)$ at $E_i = 7.7$ meV is shown in Supplementary Fig. 11. Dispersive modes indicative of spin-wave excitations are clearly observed below 7.2 K and disappear at 50 K. To estimate the energy dependence of the intensity, $S(Q, E)$ is integrated along $Q$ around the (001) reflection at 2 $Q$ ranges of 0.7 Å$^{-1}$ < $Q$ < 0.86 Å$^{-1}$ and 2.05 Å$^{-1}$ < $Q$ < 2.15 Å$^{-1}$, where contributions from nuclear and magnetic Bragg reflections are minimal. The results are shown in Supplementary Fig. 12. For the former $Q$ range, the 1.6 and 7.2 K data coincide with each other above 2 meV, whereas the 50 K data are larger than either, possibly owing to the contributions of phonons. For the latter $Q$ range, the intensity is nearly temperature independent above 2 meV, indicative of the lack of magnetic contributions there. Thus, the magnetic contributions are mostly located below 2 meV. To estimate the total magnetic scattering, we used an intensity below 2 meV.

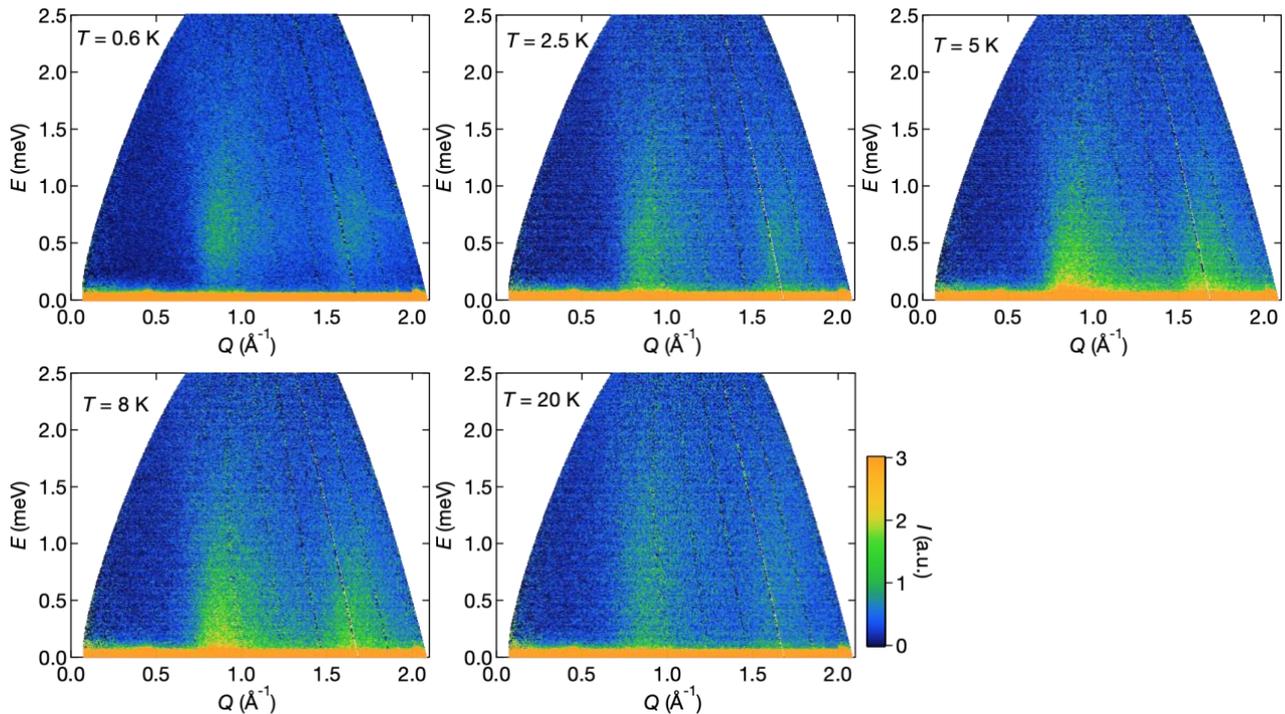

**Supplementary Fig. 8 | Powder-averaged dynamical structure factors $S(Q, E)$ of pharmacosiderite at temperatures of 0.6, 2.5, 5, 8, and 20 K and at $E_i$ = 3.1 meV.**

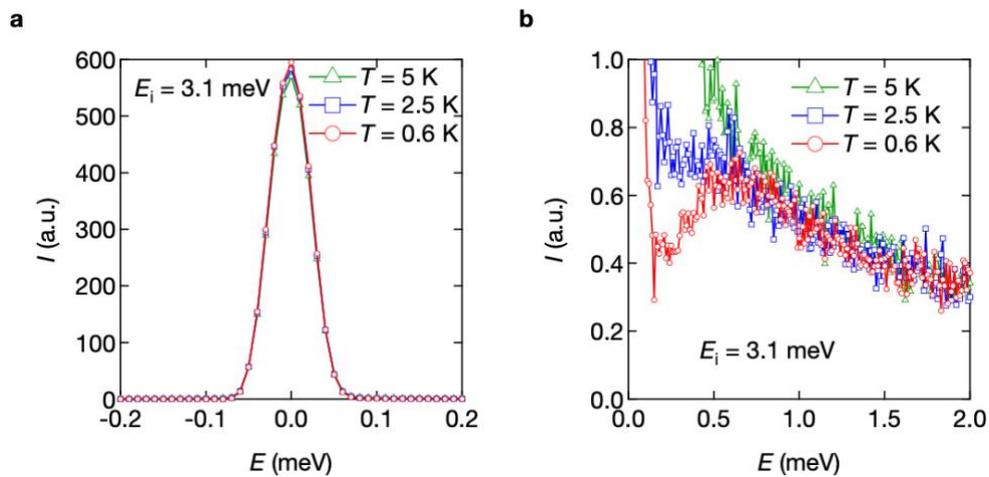

**Supplementary Fig. 9 | a. Neutron intensities obtained by integrating the $S(Q, E)$s along $Q$ around the (001) reflection (0.7 Å$^{-1}$ < $Q$ < 0.86 Å$^{-1}$) at temperatures of 0.6, 2.5, and 5 K and $E_i$ = 3.1 meV.** The low-intensity regime is expanded in **b**.

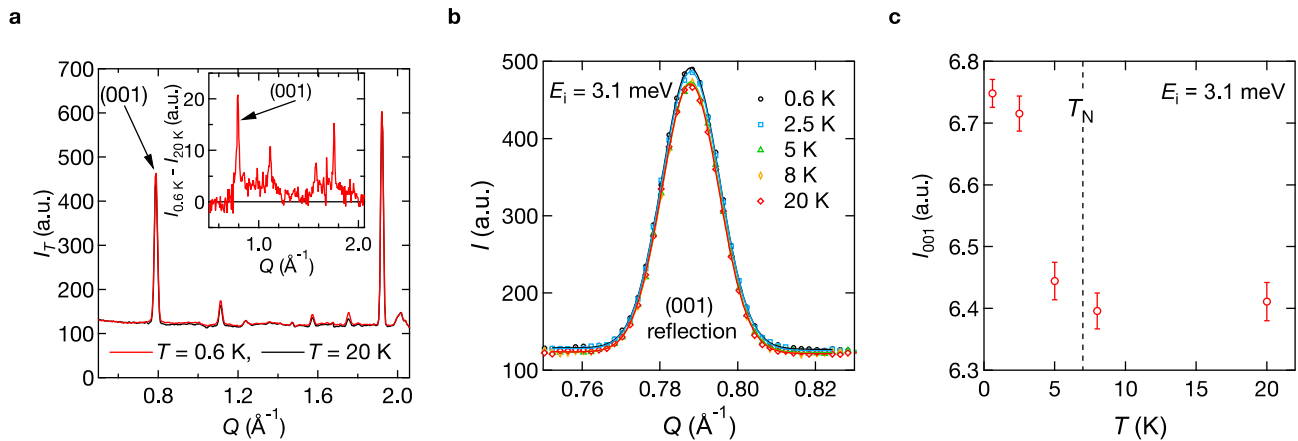

**Supplementary Fig. 10 | Estimate of elastic contribution from the integrated dynamical structure factor. a.** Diffraction patterns estimated by integrating the $S(Q, E)$ at an energy window between –0.1 and 0.1 meV at temperatures of 0.6 and 20 K. The inset shows a magnetic diffraction pattern obtained by subtracting the 20 K data from the 0.6 K data. **b.** Enlarged (001) reflection peaks at 0.6, 2.5, 5, 8, and 20 K with Gaussian fits represented by the solid lines. **c.** Temperature dependence of the intensity of the (001) peak estimated by the Gaussian fits in **b**.

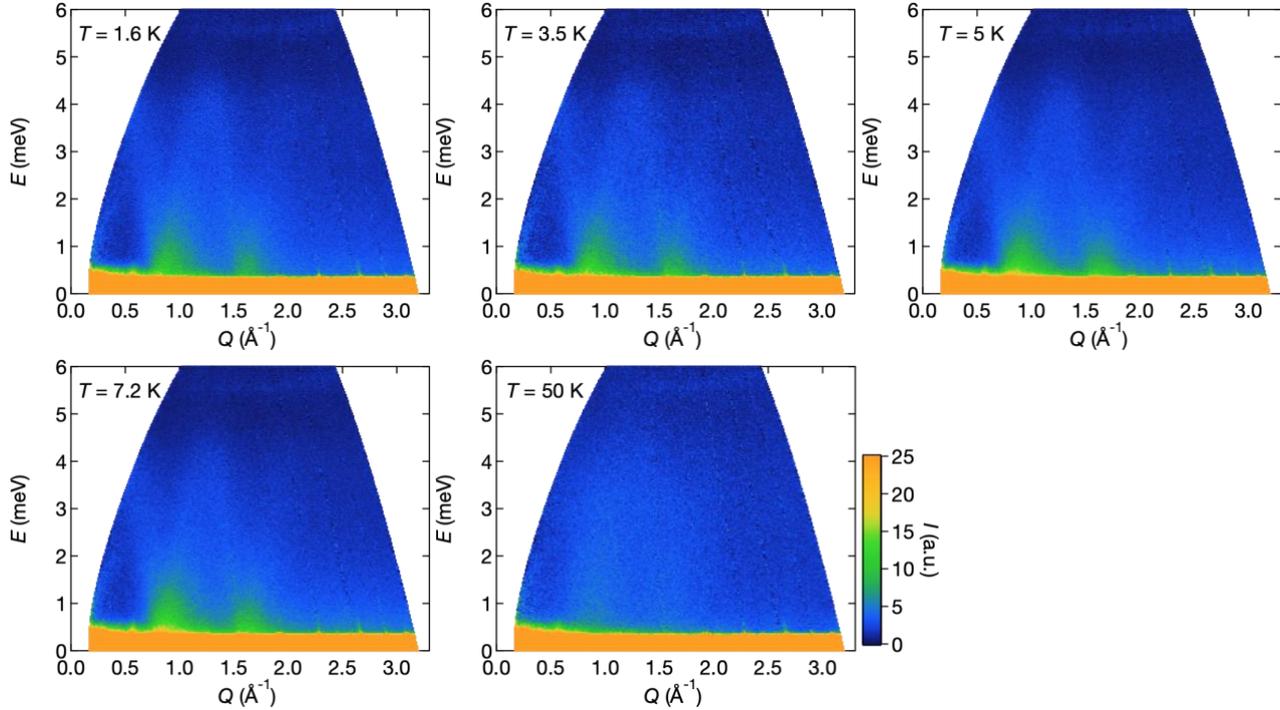

**Supplementary Fig. 11 | Powder-averaged dynamical structure factors $S(Q, E)$ of pharmacosiderite at temperatures of 1.6, 3.5, 5, 7.2, and 50 K at $E_i$ = 7.7 meV.**

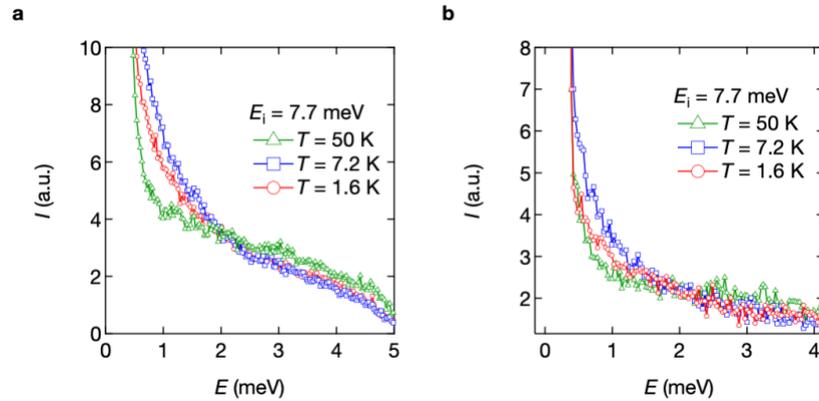

**Supplementary Fig. 12 | Estimate of magnetic contribution in the dynamical structure factor.** Neutron intensities obtained by integrating the $S(Q, E)$s along $Q$ around the (001) reflection at **a.** 0.7 Å$^{-1}$ < $Q$ < 0.86 Å$^{-1}$ and **b.** 2.05 Å$^{-1}$ < $Q$ < 2.15 Å$^{-1}$, at both of which nuclear and magnetic Bragg contributions are minimal, at temperatures of 1.6, 7.2, and 50 K and $E_i$ = 7.7 meV.

**Supplementary Note 6:    COMPARISON OF THE EXPERIMENT TO THE LINEAR SPIN WAVE THEORY**

The energy and momentum dependences of the $S(Q, E)$ of pharmacosiderite are analysed in terms of the linear spin wave theory (LSWT) using SpinW software[12]. The energy dependence of the spectra around the (001) reflection is shown in Supplementary Fig. 13a. The observed gap-like structure centred around 0.5 meV is qualitatively reproduced by the calculation, but the spectra are much sharper in the LSWT simulation. This broadening suggests that the actual magnon excitations are damped by strong fluctuations in the ordered state. In contrast, LSWT correctly captures the feature of momentum dependence in Fig. 13b, which results from two-dimensionality in the excitation, as revealed in the next section.

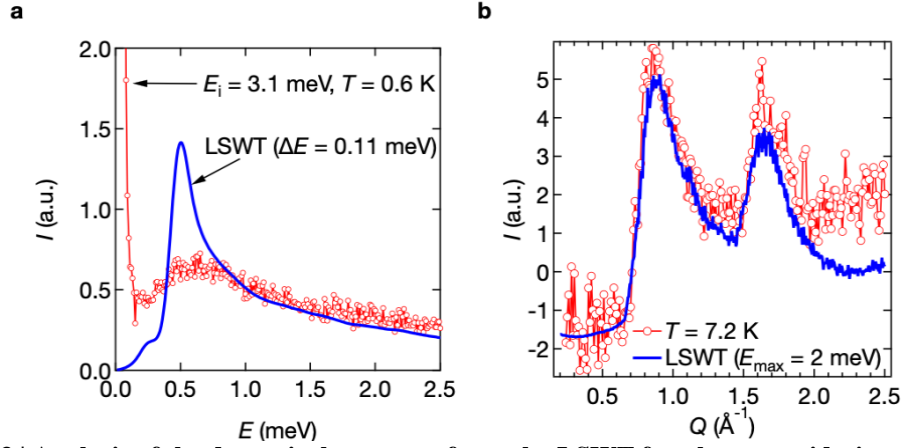

**Supplementary Fig. 13 | Analysis of the dynamical structure factor by LSWT for pharmacosiderite. a.** Energy dependence of the neutron intensity obtained by integrating the $S(Q, E)$s at $E_i = 3.1$ meV and $T = 0.6$ K over the $Q$ range of 0.7 Å$^{-1} < Q < 0.86$ Å$^{-1}$ around the (001) reflection. The solid blue line represents a simulated spectrum convoluted with the instrumental energy resolution of AMATERAS at $E_i = 3.1$ meV ($\Delta E = 0.11$ meV). The simulated spectrum is normalized to the area intensity of the experimental data at 0.14 meV $< E <$ 2.5 meV. **b**. Comparison of the momentum dependence of the $I$ at 7.2 K shown in Fig. 4e in the main text with LSWT calculation. The blue line represents a calculated intensity obtained by integrating $S(Q, E)$ at an energy range of 0 meV $< E <$ 2 meV.

## Supplementary Note 7: ANALYSIS OF THE EQUAL TIME STRUCTURE FACTOR IN TERMS OF THE LOW DIMENSIONAL SCATTERING MODELS

We consider paramagnetic scattering from classical spins in a crystal. The equal time structure factor or total scattering $I(\mathbf{Q})$ at a three-dimensional wavevector $\mathbf{Q}$ is defined as the integral of the dynamical structure factor along all available energy. $I(\mathbf{Q})$ can also be formulated as the scattering intensity from static objects as follows:

$$I(\mathbf{Q}) \propto \int_{-\infty}^{\infty} d\omega S^{\mu\mu}(\mathbf{Q}, \omega) = \sum_{\mathbf{r},\mathbf{r}'} e^{-i(\mathbf{Q}\cdot(\mathbf{r}'-\mathbf{r}))} \overline{S_{\mathbf{r}'}(0) \cdot S_{\mathbf{r}}(0)} \qquad (10).$$

Consider the powder-averaged total scattering from a $d$-dimensional short-range-ordered object. According to ref. 13, the scattering intensity is given by

$$I_d(Q) = \sum_\tau I_\tau \Phi_d(Q, \tau, D) \qquad (11).$$

Here, $Q$, $\tau$, $D$, $I_\tau$, and $\Phi_d$ represent the wavenumber, indices of the $d$-dimensional Bragg reflection, structure factor, and powder-averaged diffuse scattering from $\tau$ reflection, respectively. $D$ is regarded as the width of the Bragg peaks and is proportional to the inverse of the correlation length of the short-range order. Specifically, $\Phi_d$ in each dimension can be expressed in the following form:

$$\Phi_3(Q_0, \tau) = \frac{(2\pi)^3}{v_0 \tau^2 \sqrt{2\pi D^2}} \exp\left(-\frac{(Q_0 - \tau)^2}{2D^2}\right) \qquad (12),$$

$$\Phi_2(Q_0, \tau) = \frac{(2\pi)^2}{f_0 \sqrt{2\pi D^2}} \int_{-\infty}^{\infty} \frac{dq}{\tau^2 + q^2} \exp\left(-\frac{\left(Q_0 - \sqrt{\tau^2 + q^2}\right)^2}{2D^2}\right) \qquad (13),$$

$$\Phi_1(Q_0, \tau) = \frac{(2\pi)^{3/2}}{a_0 \sqrt{2\pi D^2}} \qquad (14).$$

Here, $v_0$, $f_0$, and $a_0$ represent the volume, surface, and length of each unit cell, respectively. To fit the $I(Q)$ of pharmacosiderite by the low-dimensional order model, the background of thermal diffuse scattering from phonons is considered as follows:

$$I_d(Q) = A + CQ^2 + \sum_\tau I_\tau \Phi_d(Q, \tau, D) \qquad (15).$$

The result of the fit using the expression above is given in Supplementary Table 5.

|  | 3D | 2D | 1D |
| --- | --- | --- | --- |
| $R_{wp}$ (%) | 6.6 | 3.3 | 6.0 |
| $I_{100}, I_{10}, I_1$ | 0.066(6) | 2.32(8) | 1.33(5) |
| $I_{011}, I_{11}$ | 0.088(8) | 1.0(1) |  |
| $I_{111}$ | 0.00(1) |  |  |
| $I_{200}, I_{20}, I_2$ | 0.076(19) | 3.9(2) | 0.96(10) |
| $I_{210}, I_{21}$ | 0.09(4) | 0.6(3) |  |
| $I_{211}$ | 0.00(3) | | |
| $I_{220}$ | 0.06(4) | 0.9(6) | |
| $I_{300}, I_{30}, I_3$ | 0.00(5) | 0.84(99) | 1.1(3) |
| $I_{310}, I_{31}$ | 0.03(3) | 0.40(98) | |
| $C$ | -2.7(4) | -1.23(16) | -0.54(17) |
| $A$ | 0.60(5) | 0.36(3) | 0.019(45) |
| $2\pi a^{-1} D^{-1}$ | 4.5(4) | 8.9(4) | 20(3) |

**Supplementary Table 5:** Parameters obtained by fitting the equal time structure factor of pharmacosiderite to the *d*-dimensional (*d* = 1, 2, and 3) short-ranged order model.

**Supplementary Note 8:  ZERO-ENERGY MODES OF SPIN WAVES IN THE COPLANAR Q = 0, $\Gamma_5$ STRUCTURE**

We discuss the instability of the **q** = 0 and $\Gamma_5$ structures in terms of the zero-energy modes in the *J*–*J'* model. The **q** = 0 and $\Gamma_5$ structures are apparently stable, as shown in Supplementary Fig. 14a, which depicts antiferromagnetic configurations between two tetrahedral spin clusters. For example, any $\Gamma_5$ order with a nonzero q value is unstable because the intercluster couplings become ferromagnetic, as shown in Supplementary Fig. 14b for neighbouring clusters related to the time-reversal operation. Interestingly, as illustrated in Supplementary Fig. 14c, an antiferromagnetic intercluster coupling similar to that in the **q** = 0, $\Gamma_5$ structure is also achieved by introducing a different irreducible representation of $\Gamma_4$.

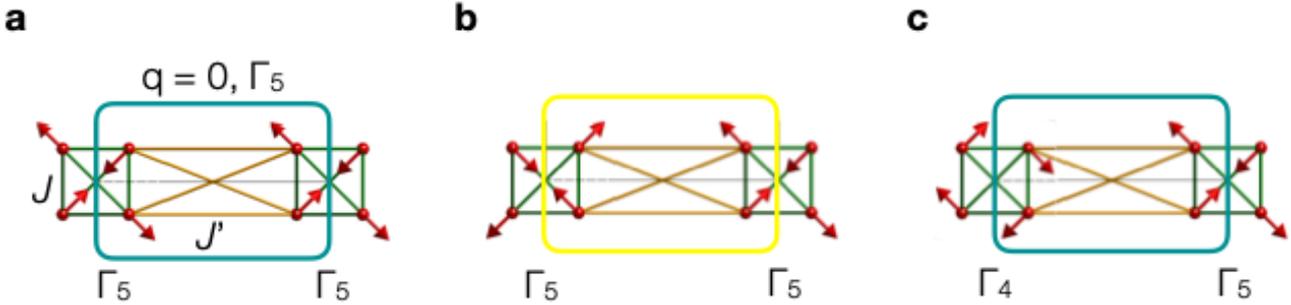

**Supplementary Fig. 14 | Comparison of the intercluster couplings. a**. **q** = 0, $\Gamma_5$ magnetic structure viewed in the direction perpendicular to the coplanar plane. An antiferromagnetic spin configuration is realized in the elongated *J'* tetrahedron. **b.** A pair of $\Gamma_5$ type clusters related by the time-reversal operation expected for **q** = (1/2, 1/2, 1/2). A nearly ferromagnetic spin configuration is realized in the elongated *J'* tetrahedron unlike **a**, which is energetically unfavorable. **c.** A pair of $\Gamma_4$ and $\Gamma_5$ type clusters. An antiferromagnetic spin configuration essentially the same as that in **a** is realized in the elongated *J'* tetrahedron.

As discussed in the main text, the **q** = 0 and $\Gamma_5$ structures possess a 2D character owing to dimensional reduction by frustration and are composed of decoupled layers perpendicular to the *c* axis (Supplementary Fig. 15). Here, we consider a 1D defect in the layer that may generate a zero-energy mode. When one $\Gamma_4$ type cluster is introduced as a point defect into the **q** = 0, $\Gamma_5$ layer, as depicted in Supplementary Fig. 15b, the neighbouring intercluster couplings become antiferromagnetic and ferromagnetic along the *a* and *b* axes with energies of $-2J'S^2$ and $+2J'S^2$, respectively. Notably, the intercluster couplings along the *c* axis remain cancelled by frustration, even in this case. On the other hand, when a string of $\Gamma_4$ clusters is created along the *b* axis instead of a single defect, as shown in Supplementary Fig. 15c, the energy-consuming ferromagnetic couplings along the *b* axis are removed and all the intercluster couplings become antiferromagnetic, so that the total energy remains the same as in the original $\Gamma_5$ structure. This means that such 1D defects can easily occur in the layer of the $\Gamma_5$ structure.

Remarkably, the spins in this 1D defect can continuously rotate without loss of energy. As illustrated in Supplementary Fig. 16, in each of the clusters of the 1D defect along the *b* axis, the $\Gamma_4$- and $\Gamma_5$-type configurations are transformed to each other by rotating the four spins cooperatively while keeping the total spin zero. There is no energy variation in the intercluster couplings to the surrounding clusters when this transformation takes place coherently in all the clusters of the 1D column. Thus, the defect can appear and disappear freely. The origin of this behaviour is the axial symmetry of the mean field from neighbouring clusters. The effective magnetic field exerted on the spins in the 1D defect from the pairs of spins in the next cluster points in the direction of *b* or –*b*, which allows a coherent "Larmor precession" around the *b* axis, as schematically depicted in Supplementary Fig. 16a. Specifically, the *ac* component of each spin in the column is rotated by an identical angle, and the *b* component is kept invariant. Therefore, a number of 1D defects are generated in the $\Gamma_5$ structure as zero-energy modes, which may disturb the LRO and lead to a unique order with a large fluctuation. However, this may not be the case at *T* = 0 because small anisotropic interactions lift the zero-energy modes to gain a finite energy gap. Nevertheless, they are stable at elevated temperatures and may be frozen by crystalline defects at lower temperatures.

It is also noted that, owing to the tetragonal symmetry of the magnetic structure, similar 1D defects made of $\Gamma_4$ type clusters can occur along the *a* axis. Then, one expects such a crossing of 1D defects, as depicted in Supplementary Fig. 16b. This is possible and does not increase the energy because one $\Gamma_5$ type cluster is generated at the cross. This situation is in contrast to the case of the weathervane modes in the kagomé antiferromagnet, in which a crossing of two 1D defects is energetically unfavourable. This means that in pharmacosiderite, a number of 1D defects are generated along both the *a* and *b* axes and effectively weaken the LRO.

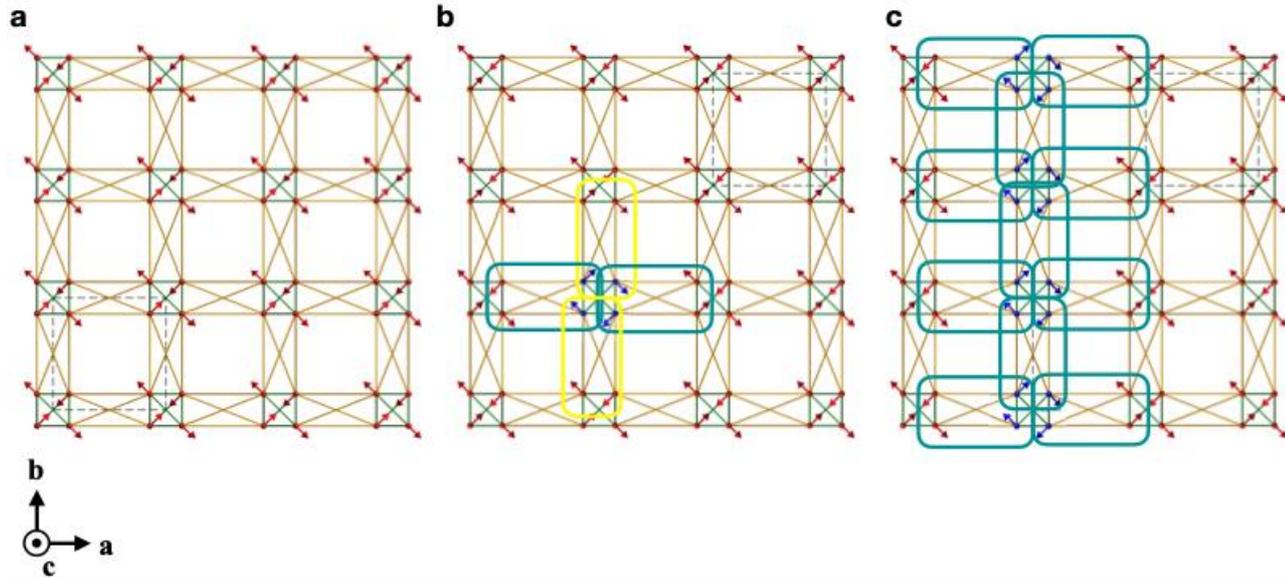

**Supplementary Fig. 15 | 2D layer of the q = 0, $\Gamma_5$ structure with $\Gamma_4$ type defect clusters. a**. q = 0, $\Gamma_5$ structure viewed along the *c* axis. Spins are aligned in the *c* plane. **b**. One of the clusters in **a** is replaced by a $\Gamma_4$ type cluster with spins shown in blue. Antiferromagnetic inter-cluster couplings are realized along the *a* axis in the blue boxes, while nearly ferromagnetic couplings occur along the *b* axis in the yellow boxes, which increases the energy. **c.** 1D defect made of $\Gamma_4$ type clusters along the *b* axis. The total energy is same as that in **a** because all the inter-cluster couplings become antiferromagnetic.

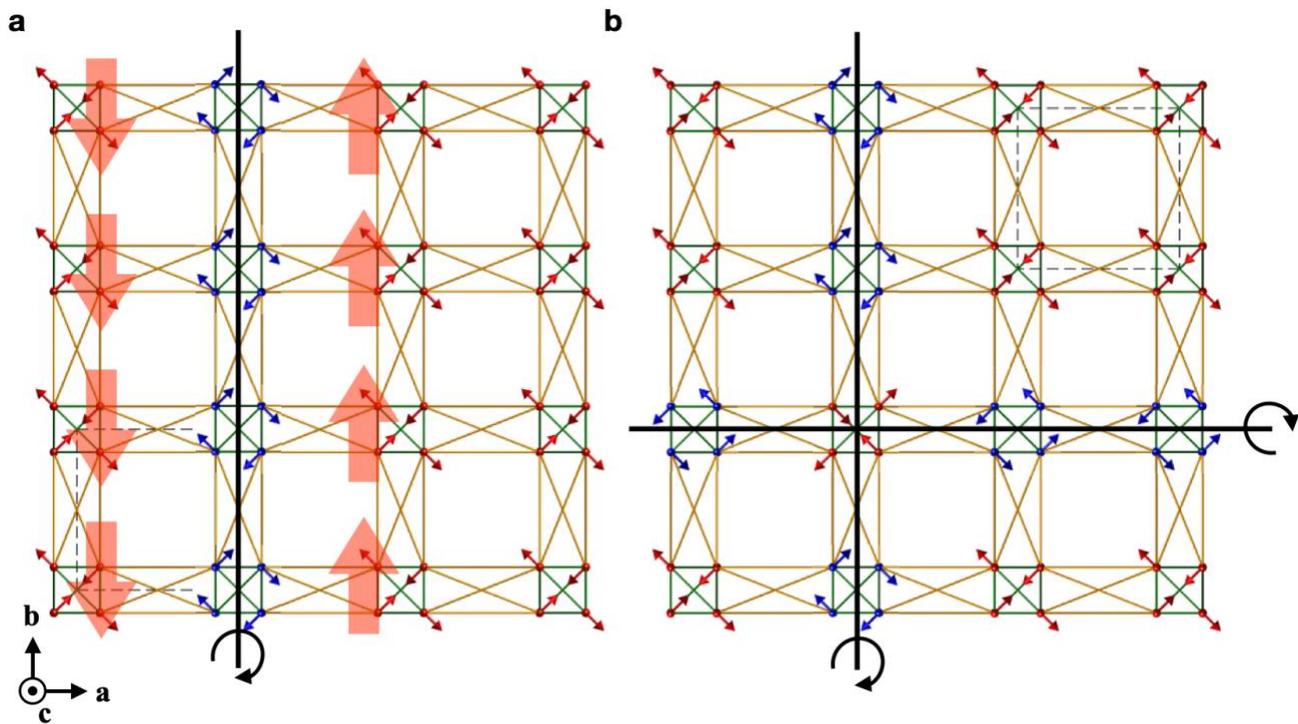

**Supplementary Fig. 16 | 1D zero-energy mode generated in the layer of the q = 0, $\Gamma_5$ structure projected along the *c* axis.** The red and blue arrows represent spins having $\Gamma_4$ and $\Gamma_5$ arrangements, respectively. **a**. Axial symmetry in the column of $\Gamma_4$ type clusters along the *b* axis. The large transparent arrow shows a total spin of two spins in a neighbouring cluster which generates an effective magnetic field along either *b* or –*b* on the spins in the defect column. This uniaxial nature allows continuous and coherent Larmor precession of all the spins in the defect column around the *b* axis without energy loss. Notably, a rotation by 180˚ leads to a transformation between the $\Gamma_4$ and $\Gamma_5$ arrangements. **b**. Intersection of two 1D defects. In addition to a 1D defect along the *b* axis, a 1D defect along the *a* axis can occur without any energy loss. At the crossing site, the $\Gamma_5$ arrangement occurs, and all the intercluster couplings remain antiferromagnetic. Thus, a number of 1D defects can be generated in the layer at *T* = 0 in the

absence of additional anisotropic interactions.

**Supplementary References**